\documentclass[a4,useAMS,usenatbib]{mn2e}

\usepackage[english]{babel} \usepackage{subfigure}
\usepackage{psfrag}
\usepackage{graphicx}

\usepackage[fleqn]{amsmath} 
\usepackage{color}

\usepackage[varg]{txfonts}

\newcommand{\noi}{\noindent}
\newcommand{\vect}{\bmath}

\newcommand{\msun}{M_\odot}

\title[Statistics of mass substructure]{Statistics of mass substructure from strong gravitational lensing: quantifying the mass fraction and mass function}

\author[S. Vegetti \& L. V. E. Koopmans]{ Simona Vegetti\thanks{E-mail:
    vegetti@astro.rug.nl} \& L. V. E. Koopmans\\ Kapteyn
    Astronomical Institute, University of Groningen, PO Box 800,
    9700\,AV Groningen, the Netherlands}
      
\begin{document}

\date{Accepted for publication in MNRAS}

\pagerange{\pageref{firstpage}--\pageref{lastpage}} \pubyear{2002}

\maketitle

\label{firstpage}

\begin{abstract}
A Bayesian statistical formalism is developed to quantify the level at which the mass-function ($dN/dm \propto m^{-\alpha}$) and the projected cumulative mass fraction ($f$) of (CDM) substructure in strong gravitational-lens galaxies, with arcs or Einstein rings, can be recovered as function of the lens-survey parameters and the detection threshold of the substructure mass. The method is applied to different sets of mock data to explore a range of observational limits: (i) the number of lens galaxies in the survey, (ii) the mass threshold, $M_{\rm{low}}$, for the detection of substructures and (iii) the uncertainty of the measured substructure masses. We explore two different priors on the mass function slope: a uniform prior and a Gaussian prior with $\alpha = 1.90 \pm 0.1$. With a substructure detection threshold $M_{\rm{low}}=3\times10^8\msun$, the number of lenses available now ($n_l=30$), a true dark-matter mass fraction in (CDM) substructure $\leq1.0\%$ and a prior of $\alpha = 1.90 \pm 0.1$, we find that the upper limit of $f$ can be constrained down to a level $\le 1.0\%$ (95\% CL). In the case of a uniform prior the complete substructure mass distribution (i.e. mass fraction and slope) can only be characterized in a number of favourable cases with a large number of detected substructures. This can be achieved by an increase of the resolution and the signal-to-noise ratio of the lensed images. In the case of a Gaussian prior on $\alpha$, instead, it is always possible to set stringent constraints on both parameters. We also find that lowering the detection threshold has the largest impact on the ability to recover $\alpha$, because of the (expected) steep mass-function slope. In the future, thanks to new surveys with telescopes, such as SKA, LSST and JDEM and follow-up telescopes with high-fidelity data, a significant increase in the number of known lenses (i.e. $\gg$$10^{4}$) will allow us to recover the satellite population in its completeness. For example, a sample of 200 lenses, equivalent in data-quality to the \emph{Sloan Lens ACS Survey} and a detection threshold of $10^8\msun$, allows one to determine $f=0.5\pm 0.1\%$ (68\% CL) and $\alpha=1.90\pm0.2$ (68\% CL).

\end{abstract}


\section{Introduction}
In the context of the cold-dark-matter paradigm, a significant number of substructures, with a steep mass function, is expected to populate the dark halo of galaxies. In galaxies as massive as the Milky Way, for example, of the order of $10^4$ substructures are predicted inside the virial radius \citep[]{Diemand08,Springel08}, although only about $20$ have been so far observed \citep[]{Zucker04,Willman05,Belokurov06,Grillmair06,Martin06,Sakamoto06,Zucker06b, Zucker06a,Belokurov07b,Ibata07, Irwin07,Majewski07,Walsh07,Zucker07,Belokurov08}. A clear comparison between the simulated and the physical reality, however, is strongly hampered by the difficulty of directly observing substructures in distant galaxies, as well as in the Local Group.\\
\noi While major improvements in the observations and numerical simulations have not yet significantly alleviated the satellite crisis, new techniques have been proposed for the indirect and direct detection of subhaloes that may have a high mass-to-light ratio. In our own Galaxy, CDM substructures can be, in principle, identified via their effect on stellar streams \citep[]{Ibata02, Mayer02} or via the dark matter annihilation signal from their high-density centres \citep[]{Bergstrom99, Calcaneo00, Stoehr03,Colafrancesco06}; gravitational lensing, on the other hand, allows for direct detection (measurement of the substructure gravitational signature) in the central regions of galaxies through flux-ratio anomalies and distortions of extended Einstein rings and arcs. \citep[e.g.][]{Mao98,Metcalf01,Dalal02, Koopmans05}. Interestingly enough, results based on flux-ratio anomalies reverse the satellite crisis with a recovered mass fraction in substructure which seems to be larger than predicted by numerical simulations \citep[e.g.][]{Mao04, Maccio06}. While this discrepancy might not be easily accommodated by an increase in the resolution of simulations, the correct interpretation of the flux-ratio anomalies is still subject of discussion.\\ 
\noi In \citet{Vegetti09}, we introduced a new adaptive-grid method, based on a Bayesian analysis of the surface brightness distribution of highly magnified Einstein rings and arcs, that allows the identification and precise quantification of substructure in single lens galaxies. This technique does not depend on the nature of dark matter, on the shape of the main galaxy halo, strongly on the density profile of the substructure, nor on the dynamical state of the system. It can be applied to local galaxies as well as to high redshift ones, as long as the lensed images are highly magnified, extended and have a high signal-to-noise ratio. Unlike flux ratio anomalies, extended optical images are little affected by differential scattering in the radio or microlensing in the optical and X-ray. If substructures are located close to the lensed images, the method allows the determination of both their mass and position, although as the distance between the substructure and the Einstein ring increases, the mass model becomes more degenerate. Finally, thanks to its Bayesian framework, the method of \citet{Vegetti09} requires hardly any subjective intervention into the modelling and any assumption can be objectively tested through the Bayesian Evidence \citep[]{MacKay92}. \\ 
\noi In this paper, we show how results from data sets with more than one lens system (i.e. the number of detections and their masses) can be combined to statistically constrain the fraction of dark matter in substructure and their mass function, as function of the survey parameters and limits. The combination of multiple data sets becomes important when trying to constrain the slope of the substructure mass function. In a single lens potential, in fact, the number of detectable substructures can, in certain cases, be as small as zero or one. More than one lens is, therefore, required in order to improve statistics and properly \emph{sample} the mass function. On the contrary an upper limit to the mass fraction in substructure can always be set.\\ 
The paper is organized as follows. In Section 2, we outline the statistical formalism of satellite detection; in Section 3, application of the method is discussed and in Section 4 conclusions are drawn.

\section{Bayesian interpretation of substructure detections}
In this section, we outline a Bayesian formalism that allows us to give a statistical interpretation to the detection of dark substructure in gravitational lens galaxies and to recover the properties of the substructure population. Thanks to Bayes' theorem, the likelihood of measuring a mass substructure, can be translated into probability density distributions for the substructure mass fraction $f$ and their mass function $dN/dm\propto m^{-\alpha}$ (when $dN/dm$ is normalised to unity, we refer to it as $dP/dm$), as function of the mass measurement errors and the model parameters. The latter include the minimum and maximum mass of dark matter substructure, $M_{\rm{min}}$ and $M_{\rm{max}}$, between which the mass fraction is defined, and the lowest and highest mass we can detect, $M_{\rm{low}}$ and $M_{\rm{high}}$. More precisely, $M_{\rm{low}}$ should not be interpreted as a hard detection threshold, but as a statistical limit above which we believe a detection to be significant, at the level set by the mass measurement error $\sigma_m$ (i.e. with a signal-to-noise ratio of $M_{\rm{low}}/\sigma_m$).

\subsection{Likelihood of the substructure measurements}
We derive an expression for the Likelihood of observing $n_s$ substructures for a given sample of $n_l$ lenses. 
We assume the cumulative dark-matter mass fraction of substructure $f(<R/R_{\rm{vir}})$, within a cylinder of projected radius $R$ of the lens, to be the same for all lens galaxies and the detections of multiple substructures in one galaxy to be independent from one another.
We also assume that the number of substructures populating a given galactic halo fluctuates with a distribution which is Poissonian. We know that not all these substructures are observable, but only those with the right combination of mass and position.\\ 
\noi The Likelihood of measuring $n_s$ substructures, each of mass $m_i$, in a single galaxy, in general can be expressed as the probability density of having $n_s$ substructures in the considered lens, times the normalised probability density $P\left(m_i,R~|~\vect{p},\alpha\right)$ of actually observing the mass $m_i$ within the projected radius $R$, where $\vect{p}=\left\{M_{\rm{min}},M_{\rm{max}},M_{\rm{low}},M_{\rm{high}}\right\}$ is a vector containing all the fixed model parameters introduced above, so that
\begin{equation}
{\cal L}\left( n_s,\vect{m}~|~\alpha,f,\vect{p}\right) = \frac{e^{-\mu(\alpha,f,< R)}~{\mu(\alpha,f,< R)}^{n_s}}{n_s!}\prod_{i=1}^{n_s}P\left(m_i,R~|~\vect{p},\alpha\right) \,,
\label{equ:single_lens}
\end{equation}
\noi where $\vect{m}$ contains all the substructure masses $m_i$ and $\mu(\alpha,f,<R)$ is the expectation value of the number of substructures in a generic aperture with dark matter mass $M_{\rm{DM}}(<R)$, which will be discussed in more details in the following section. The second factor in equation (\ref{equ:single_lens}) describes the likelihood with which a substructure can be identified as a function of its mass $m_i$ and position $R$. Although the specific shape of $P\left(m_i,R~|~\vect{p},\alpha\right)$ might change from one system to another, its overall trend is essentially the same for all lenses: high mass perturbations, on or close to the lensed images are the most likely to be confidently observed. For each considered lens system, $P\left(m_i,R~|~\vect{p},\alpha\right)$ can be reconstructed by Monte Carlo explorations of the model parameter space \citep{Vegetti09}.\\
Here, for the sake of simplicity, we assume that the probability of measuring a mass $m_i$ for a single substructure is a Gaussian function independent of position. This is indeed true for small regions around the Einstein ring of the lensed images, so that $P\left(m_i,R~|~\vect{p},\alpha\right)$ reduces to 
\begin{equation}
P\left(m_i~|~\vect{p},\alpha\right) = \frac{\int_{M_{\rm{min}}}^{M_{\rm{max}}}{\left. \frac{dP}{dm}\right|_{true}\frac{e^{-\left(m-m_i\right)^2/2{\sigma_m}^2}}{\sqrt{2\pi}\sigma_m} dm}}{\int_{M_{\rm{low}}} ^{M_{\rm{high}}} { \int_{M_{\rm{min}}} ^{M_{\rm{max}}} {\left.\frac{dP}{dm}\right|_{true}}\frac{e^{-\left(m-m^\prime\right)^2/2{\sigma_m}^2}}{\sqrt{2\pi}\sigma_m} ~dm~dm^\prime}}\,,
\end{equation}
\noi where the Gaussian convolution expresses the scatter in the substructure mass due to a measurement uncertainty and $dP/dm|_{true}$ is the true substructure mass function as defined in equation (6). The assumption of Gaussian errors on the substructure mass is motivated by results from \citet{Vegetti09}.\\
\noi Equation (\ref{equ:single_lens}) can be easily extended to the case in which more than one lens is considered. Because the identification of a substructure in one system does not influence what we infer about another satellite in another lens, the Likelihood function for a set of $n_l$ lenses is simply the product of the independent likelihoods of individual detections, 
\begin{equation}
{\cal L}\left( \{n_s,\vect{m}\}~|~\alpha,f,\vect{p} \right) = \prod_{k=1}^{n_l}{\cal L}\left( n_{s,k},\vect{m}_k~|~\alpha,f,\vect{p} \right)\,.
\label{equ:multiple_lens}
\end{equation}  
\noi We now go through the details of the expectation value $\mu(\alpha,f,<R)$, characterising the Poisson distribution of the number of substructures in the potential of a generic lens galaxy.

\subsection{Substructure expectation value}
In the ideal case of infinite sensitivity to a mass perturbation, one would be able to recover the full mass function between $M_{\rm{min}}$ and $M_{\rm{max}}$. In practice, only substructures with mass between $M_{\rm{low}}$ and $M_{\rm{high}}$ are observable, hence the expectation value for observable substructures is
\begin{equation}
	\mu(\alpha,f,<R)= \mu_0(\alpha,f,<R,\vect{p}) \int_{M_{\rm{low}}} ^{M_{\rm{high}}} { \left.\frac{dP}{dm}\right | _{true}~dm}
	\label{equ:mu}	
\end{equation}
\noi with the expectation from the full mass function given by
\begin{multline}
	\mu_0(\alpha,f,<R,\vect{p}) = \frac{ f(<R)~ M_{\rm{DM}}(<R)} {\int_{M_{\rm{min}}}^{M_{\rm{max}}}{m~\left.\frac{dP}{dm}\right |_{true}~dm}}=\\ \\ =f(<R)~M_{\rm{DM}}(<R)\left\{
	\begin{array}{ccc}
	\frac{ \left(2 -\alpha\right)~ \left(M_{\rm{max}}^{1-\alpha}~-~M_{\rm{min}}^{1 -\alpha}\right)} {\left(1 -\alpha \right)~ \left(M_{\rm{max}}^{2 -\alpha}~-~M_{\rm{min}}^{2 -\alpha}\right)} & \alpha \neq 2 & ,~ \alpha \neq 1\\
	\\
	-\frac{\left(M_{\rm{max}}^{-1}~-~M_{\rm{min}}^{-1}\right)} {\log\left(M_{\rm{max}}~/~M_{\rm{min}}\right)} & \alpha = 2\\
	\\
	\frac{\log{\left(M_{\rm{max}} / M_{\rm{min}}\right)}} { \left(M_{\rm{max}}~-~M_{\rm{min}}\right)} & \alpha =1 &\\
	\end{array}
	\right.
\label{equ:mu_not}	
\end{multline}
\noi Hence, we assume the normalised true mass function to be given by a power-law
\begin{equation}
\left.\frac{dP}{dm}\right|_{true}=\left\{
 \begin{array}{cc}
 \frac{\left(1-\alpha\right)~m^{-\alpha}}{ \left(M_{\rm{max}}^{1-\alpha}~-~M_{\rm{min}}^{1 -\alpha}\right)}& \alpha \neq 1\\
\\
 \frac{m^{-\alpha}}{ \log{\left(M_{\rm{max}}/M_{\rm{min}}\right)}}& \alpha = 1 
\end{array}
	\right.
\end{equation}
\noi $ M_{\rm{DM}}(<R)$ and $f(<R)$ are the cumulative mass in dark matter and the cumulative fraction of dark matter in subhaloes within the considered radius, respectively.
However, the presence of noise on the data and the statistical uncertainty with which masses are measured introduce a scatter in the observed mass function, so that detections can be spread inside or outside our observational limits. The significance of this effect depends on the substructure mass, with lower masses being affected by a larger relative uncertainty. The observed mass function $dP/dm|_{conv}$ can then be written as a convolution of the true mass function with the error distribution, which we assume to be Gaussian, hence
\begin{multline}
\mu(\alpha,f,<R,\vect{p}) = \mu_0(\alpha,f,<R,\vect{p}) \int_{M_{\rm{low}}} ^{M_{\rm{high}}} { \left. \frac{dP}{dm}\right |_{conv}~dm} = \\ \mu_0(\alpha,f,<R,\vect{p}) \int_{M_{\rm{low}}} ^{M_{\rm{high}}} { \int_{M_{\rm{min}}} ^{M_{\rm{max}}} {\left.\frac{dP}{dm}\right |_{true}}\frac{e^{-\left(m-m^\prime\right)^2/2{\sigma^2_m}}}{\sqrt{2\pi}\sigma_m} ~dm~dm^\prime}\,.
\label{equ:mu_conv}
\end{multline}
\noi A mathematical proof for this procedure can be found in the Appendix.

\subsection{Posterior probability function of $\alpha$ and $f$} 
Given a set of observations, in which a certain number of substructures are identified and their masses are quantified for one or more lenses, equations (\ref{equ:single_lens}) and (\ref{equ:multiple_lens}) can be used to infer the mass fraction and the mass function of the underlying subhalo population. Bayes' theorem relates the Likelihood function of the observations to the joint posterior probability of $\alpha$ and $f$ in the following way
\begin{equation}
P\left( \alpha,f ~|~ \{n_s,\vect{m}\},\vect{p} \right) =\frac{{\cal L}\left( \{n_s,\vect{m}\}~|~\alpha,f,\vect{p}\right)P\left( \alpha,f ~| ~\vect{p}\right)}{P\left( \{n_s,\vect{m}\}~|~\vect{p}\right)}\,,
\end{equation}
\noi where $P\left( \alpha,f ~| ~\vect{p}\right)$ is the prior probability density distribution function of $\alpha$ and $f$. For the mass fraction we assume a non-informative uniform prior between the limits $f_{\rm{min}}=0$ and $f_{\rm{max}}=1$; while we test two different priors for $\alpha$. We assume in one case a uniform distribution between $\alpha_{min}=1.0\,$ and $\alpha_{max}=3.0$ and in the other case a Gaussian function with centre on 1.90 and a standard deviation of 0.1, as found in almost all numerical simulations. We refer to the next section for a more detailed description of the prior probability density distributions. 

\subsection{Dark matter mass}
As shown in equation (\ref{equ:mu_not}), the number of substructures expected in a given potential is a function of the dark mass $M_{\rm{DM}}(<R)$. In this section we show how this mass can be empirically estimated. Specifically, we are interested in the cumulative mass within a narrow annulus $\Delta R= 2~\delta R=0.6''$ centred around the Einstein radius $R_{\rm{E}}$, where the formalism introduced can be considered valid. In the approximation of a small annulus, the dark matter mass contained in it can be approximated as 
\begin{equation}
M_{\rm{DM}}\approx 4\pi~ R_{\rm{E}}~\Sigma_{\rm{DM}}(R_{\rm{E}})~\delta R=4\pi~ R_{\rm{E}}~\left[1-\frac{ \Sigma_{\rm{*}} }{ \Sigma_{\rm{tot}} }\right] ~\Sigma_{\rm{tot}}~\delta R\,.
\end{equation}
\noi $\Sigma_{\rm{tot}}$ and $\Sigma_{\rm{*}}$ are the total and the stellar projected mass density, respectively.
It has been found by many authors, that the total mass density has a profile which is close to isothermal \citep[e.g][]{Gerhard01,Koopmans02, Koopmans06, Czoske08, Koopmans209}. Similarly, a \citet[]{Jaffe83} profile approximates the stellar mass distribution well, i.e., 
\begin{equation}
\Sigma_{\rm{tot}}(R)=\frac{\Sigma_c~R_{\rm E}}{2R}
\end{equation}
\noi and
\begin{equation}
\Sigma_{\rm{*}}(R)= C \left\{\frac{1}{4~\tilde{r}}+\frac{1}{2\pi}\left[ \left(1- \tilde{r}^2\right)^{-1}-\right.\right.\left.\left.\left(1- \tilde{r}^2\right)^{-3/2}\left(2- \tilde{r}^2\right)\cosh^{-1}(~\tilde{r}^{-1})\right]\right\}\,
\end{equation}
\noi with $\tilde{r}=R/r_s$.\\
\noi $\Sigma_c$ is the critical surface density for lensing and $r_s$ is the scaling radius of the Jaffe profile, which relates to the effective radius of the lens galaxy as $r_s= R_e/0.74$. We have assumed in the above equations that the Einstein radius and the effective radius are related to each other by $R_e=2R_{\rm E}$, which is approximately the case for the average SLACS lens \citep[]{Bolton08}. Obviously, this is not exactly true for any of these lenses, but we are not interested in analysing an exact reproduction of the SLACS sample but just an average realisation of it. This can also be considered as a fair realisation of a typical massive early-type galaxy, given that both the internal and environmental properties of the SLACS lenses do not significantly depart from those of other early-type galaxies with comparable velocity dispersion and baryonic properties \citep{Bolton06,Treu09}. This assumption is in any case justified and does not influence our results; in fact, only the cumulative dark matter mass that is probed by all lenses is of relevance and its average value is not altered by the assumption of $R_e=2R_{\rm E}$.\\
\noi The normalization constant $C= 0.74 \Sigma_{c}$ can be derived by imposing $\Sigma_{\rm{tot}}\rightarrow\Sigma_{\rm{*}}\approx \frac{C r_s }{ 4R}$ for $R\rightarrow 0$, that is by imposing that asymptotically for $R\rightarrow 0$ the mass density becomes that of stars only, i.e. the \emph{maximum bulge} assumption.\\
\noi Because we assume that $R_e/r_s$ is a constant and $R_e/R_E$ is also a constant, on average the projected DM mass fraction within an annulus of $2\delta R$ around the Einstein radius is a constant with a value of about $63\%$. Hence we find that on average $\Sigma_{\rm{DM}} \approx 0.5 \times 0.63 ~\Sigma_c$ at the Einstein radius for all lenses. This fraction is consistent with other observations \citep[e.g.][]{Gavazzi08,Schechter02}. The number of substructures thus becomes only a function of the size of the aperture and the critical density, which itself only weakly depends on the source and lens redshifts. \\

\section{Data realisation and analysis}
In this section, we use a series of mock data sets to show that the formalism presented here allows us to constrain, in different situations, the properties of the substructure population.
Different data sets of lens galaxies with $n_l=$ 10, 30 or 200 are analysed. A sample based on the masses and radii of the SLACS lenses, ranked from the highest to the lowest mass enclosed within $R_{\rm{E}}$, is used to construct the dark matter mass function, while results from the method in \citet{Vegetti09} are used to set the observational limits and the substructure mass uncertainty. In each lens, the substructures are distributed in mass according to $dP/dm|_{true}$ with $M_{\rm{min}}=4.0\times10^6{\msun}$ and $M_{\rm{max}}=4.0\times10^9{\msun}$ \citep{Diemand07b,Diemand07a} and fluctuate in number with a Poisson probability distribution of expectation value $\mu(\alpha,f,<R)$, given by equation (\ref{equ:mu}). \\
\noi Although it was shown by \citet{Vegetti09} that the detection probability is a joint function of the substructure mass and position, this probability can be assumed to be independent of the perturber's position if a sufficiently small annulus around the Einstein ring is considered, in which case the equations presented in the previous sections hold. We consider a region of $\pm 0.3$ arcsec around $R_{\rm{E}}$, over which a typical SLACS lens shows a reasonable surface brightness of its images and within which we might expect to detect CDM substructures using the method of \citet{Vegetti09}. The extent of this area is, also, such that the mass fraction of substructures can be considered constant in radius, hence $f(R)=\rm{const.}$ over $\Delta R$.\\
\noi Each mock data set is characterized by a different fraction $f_{\rm{true}}$ of substructures, while the true slope of the mass function is kept fixed at $\alpha_{\rm{true}}=1.90$, as suggested by numerical simulations. Results from the latest high-resolution numerical simulations seem to indicate a dark matter mass fraction in substructures within a cylinder of projected 10 kpc which is between 0.3\%\footnote{This value has been obtained by including only particles within $R_{{200}_{crit}}$, adding masses of all those particles that are in subhaloes within the considered cylinder and considering the median over 100 projected directions. No extrapolation beyond the resolution limit or cut on the subhalo mass is involved.} (Mark Vogelsberger, private communication) and 0.5\% \citep[]{Diemand08}. We therefore discuss three different cases $f_{\rm{true}}=0.1,0.5,2.5\%$. The latter high fraction is included because it is close to that suggested by the median value inferred from flux-ratio anomaly studies \citep{Dalal02}.

\subsection{Observational limits on the substructure mass}
We explore the effect of different values of the lowest detectable mass $M_{\rm{low}}$, which is set at the statistical threshold above which we are confident that other effects do not create too many false events. 
We have shown in \citet{Vegetti09} that, given the current HST data quality of the SLACS lenses, a lower mass limit for a significant detection can be set around $10^8{\msun}$, depending on how close the perturbers are located with respect to the lensed images and the structure of the lensed images. However, these limits have been determined for cases not affected by systematic errors; we adopt $M_{\rm{low}}=0.3,1.0,3.0 \times10^{8}\msun$. We set a finite upper mass limit $M_{\rm{high}}=M_{\rm{max}}$. 

\subsection{Priors on $\alpha$ and $f$}
While the mass fraction of satellites is the most uncertain parameter, most studies seem to agree on the mass function, with values of the slope $\alpha$ ranging from 1.8 to 2.0 \citep[e.g.][]{Helmi02,Gao04,Diemand08, Springel08}. This is a direct consequence of the assumed cold nature of the dark matter particles. We analyse two scenarios: the first, relying on results from numerical simulations, assumes for $\alpha$ a Gaussian prior centred at $\alpha_{\rm{true}}$ with standard deviation $\sigma_{\alpha}=0.1$; the second scenario, allowing for more freedom, has a uniform prior between 1.0 and 3.0. In general, the first case can be seen as a test of N-body simulations and the second as a test of nature itself, although in the specific case of this paper the data have been created with a combination of fraction and slope typical of a standard cosmology. The former prior, obviously, provides a well-defined mass-function slope, but it also reduces the uncertainty in the mass fraction, which is less well-defined than $\alpha$ and can vary considerably between similar simulations. Different phenomena can affect the mass fraction, as for example the resolution of the simulations, or the lack, in high resolution simulations, of gas physics, that could sensibly influence the substructure survival (i.e. $f$ could be even higher than what the simulations suggest in the inner regions of the galactic haloes). We assume, conservatively, for $f$ a uniform prior ranging between $0\%$ and $100\%$.
\subsection{Results in the limit of no mass measurement errors}
\begin{figure*}
\begin{center}
\begin{minipage}[b]{0.33\linewidth}
\includegraphics[scale=0.25]{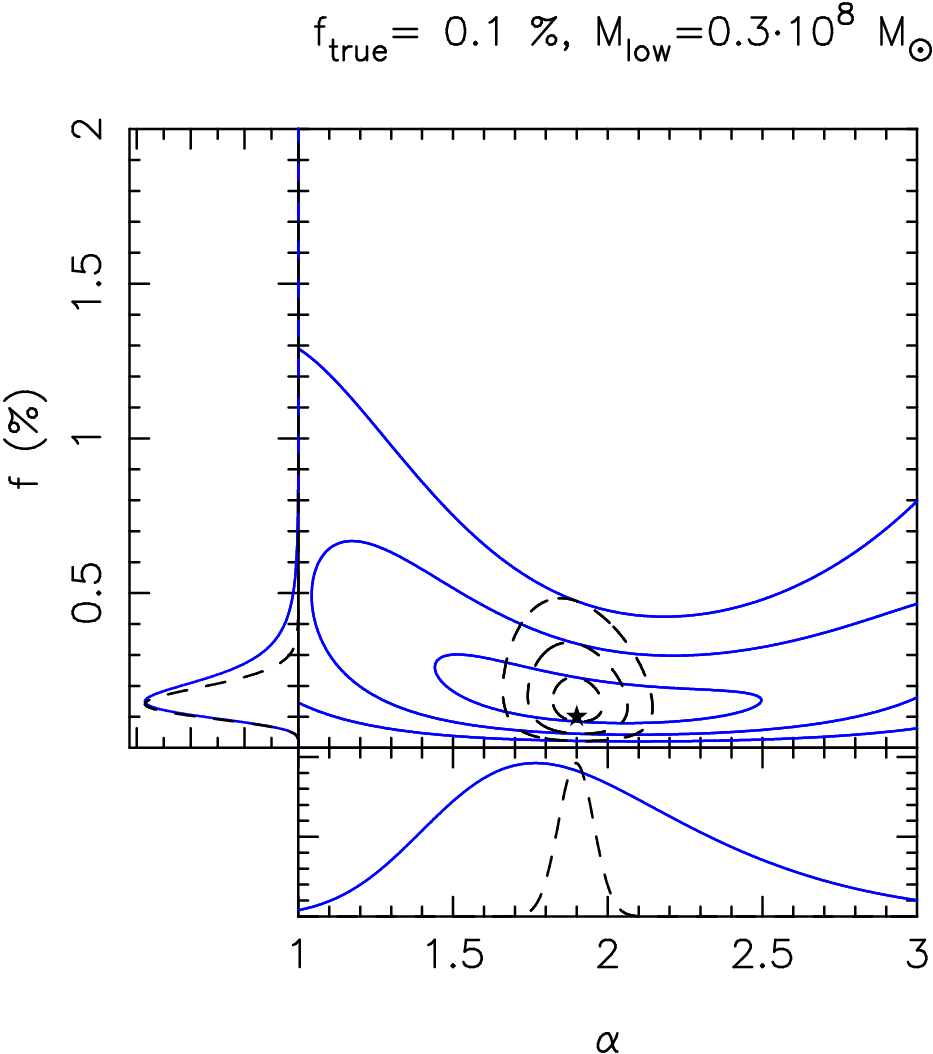}
\end{minipage}
\hspace{0.1cm}
\begin{minipage}[b]{0.33\linewidth}
\includegraphics[scale=0.25]{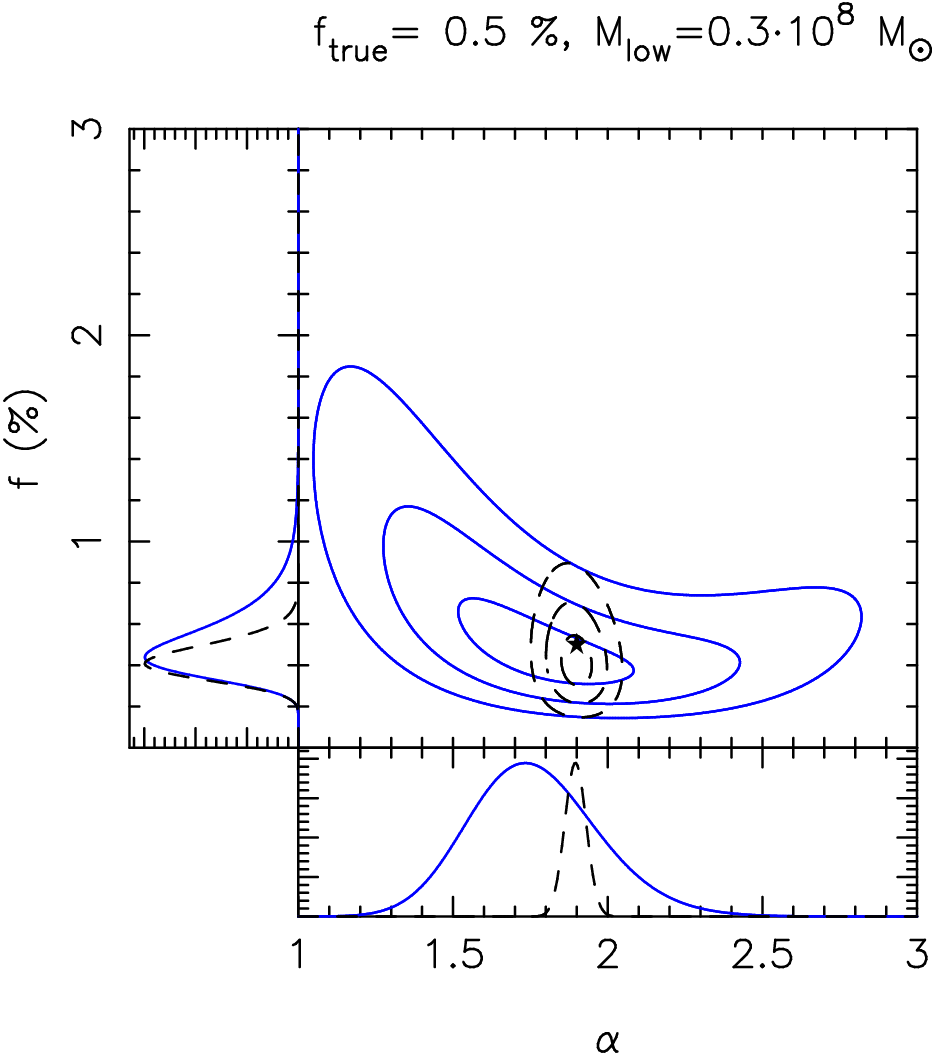}
\end{minipage}
\hspace{0.01cm}
\begin{minipage}[b]{0.25\linewidth}
\includegraphics[scale=0.25]{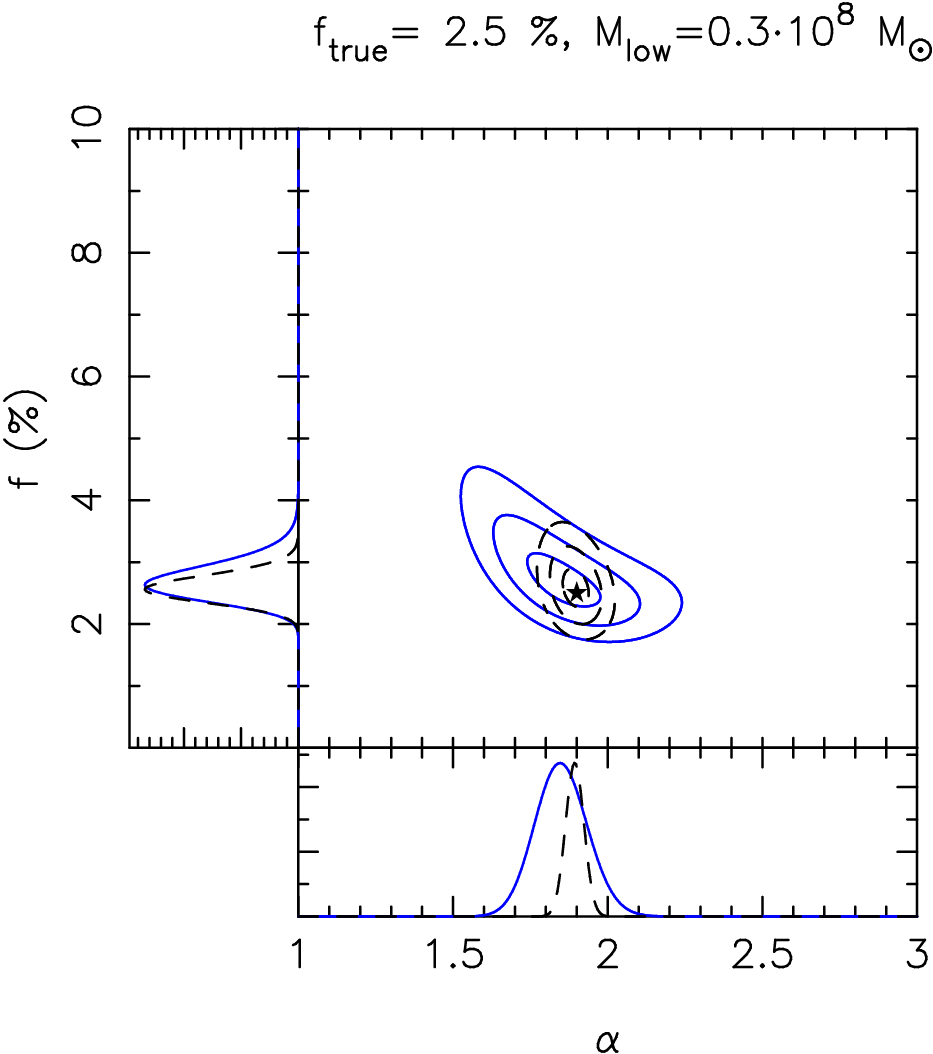}
\end{minipage}

\begin{minipage}[b]{0.33\linewidth}
\includegraphics[scale=0.25]{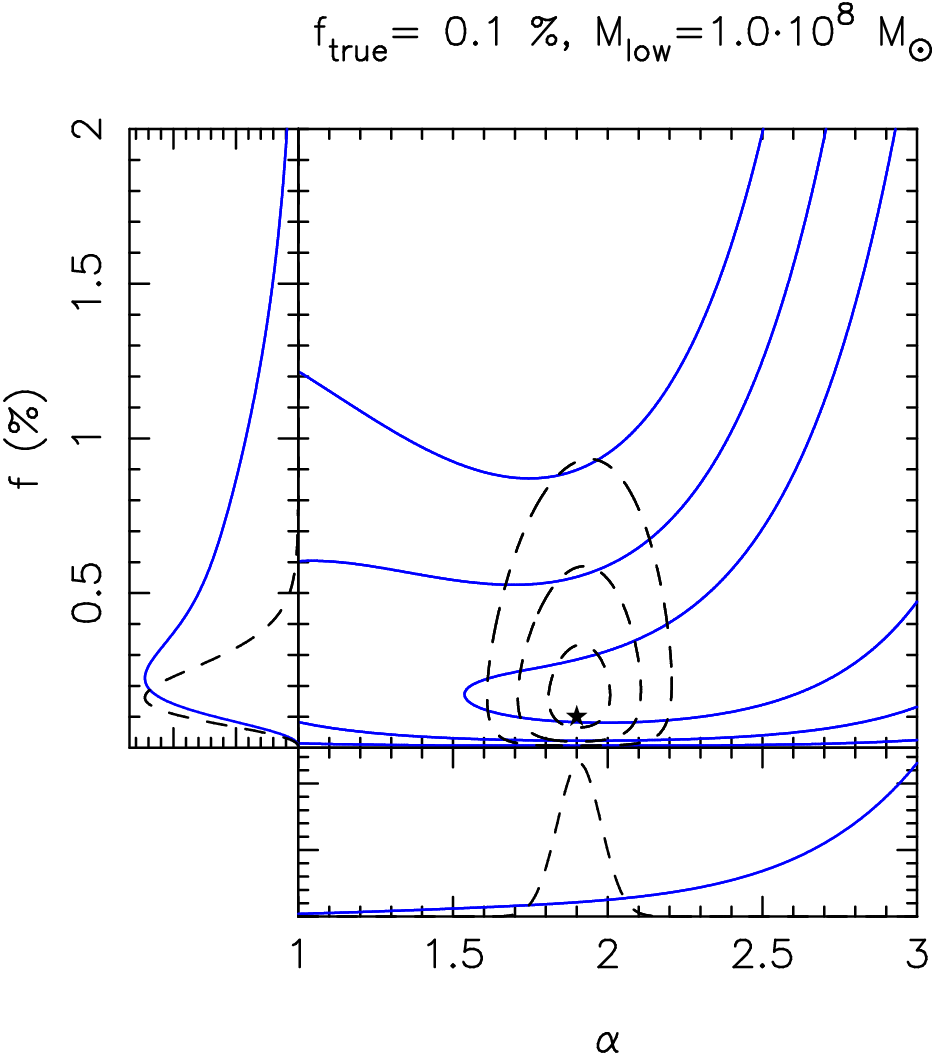}
\end{minipage}
\hspace{0.1cm}
\begin{minipage}[b]{0.33\linewidth}
\includegraphics[scale=0.25]{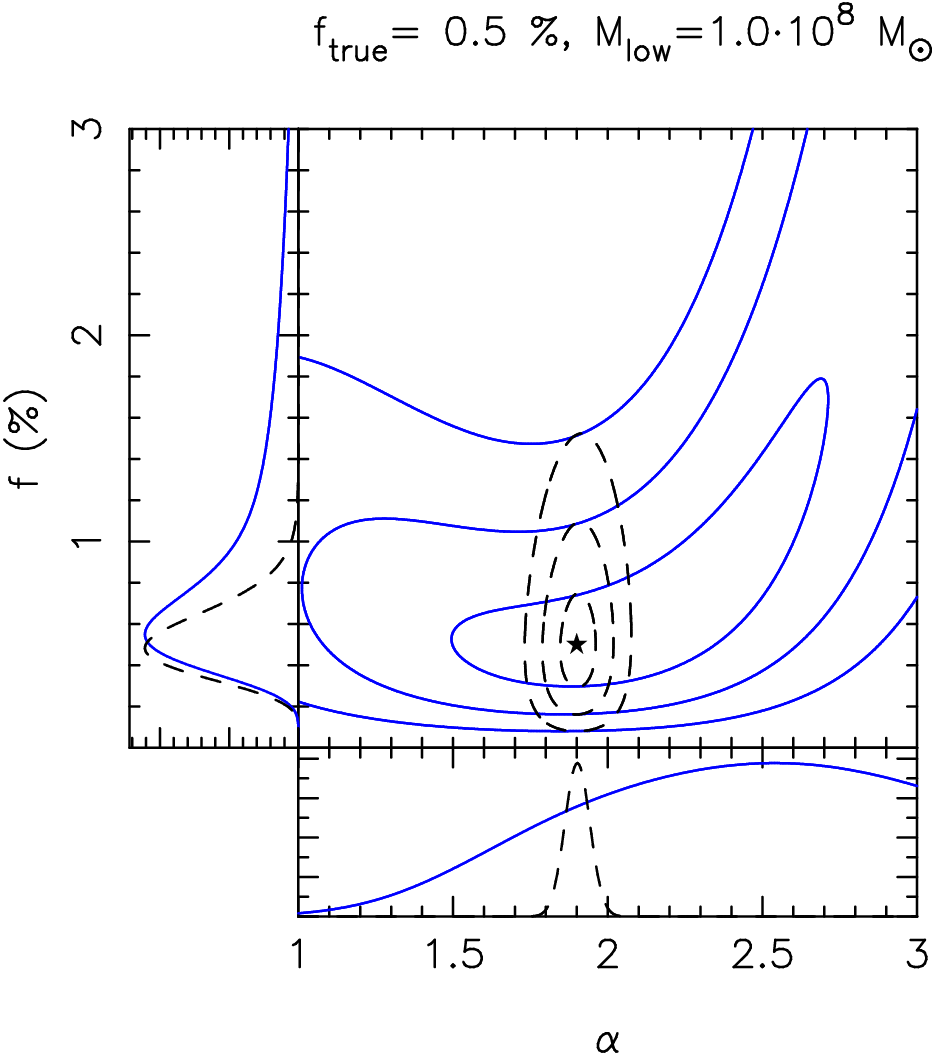}
\end{minipage}
\hspace{0.01cm}
\begin{minipage}[b]{0.25\linewidth}
\includegraphics[scale=0.25]{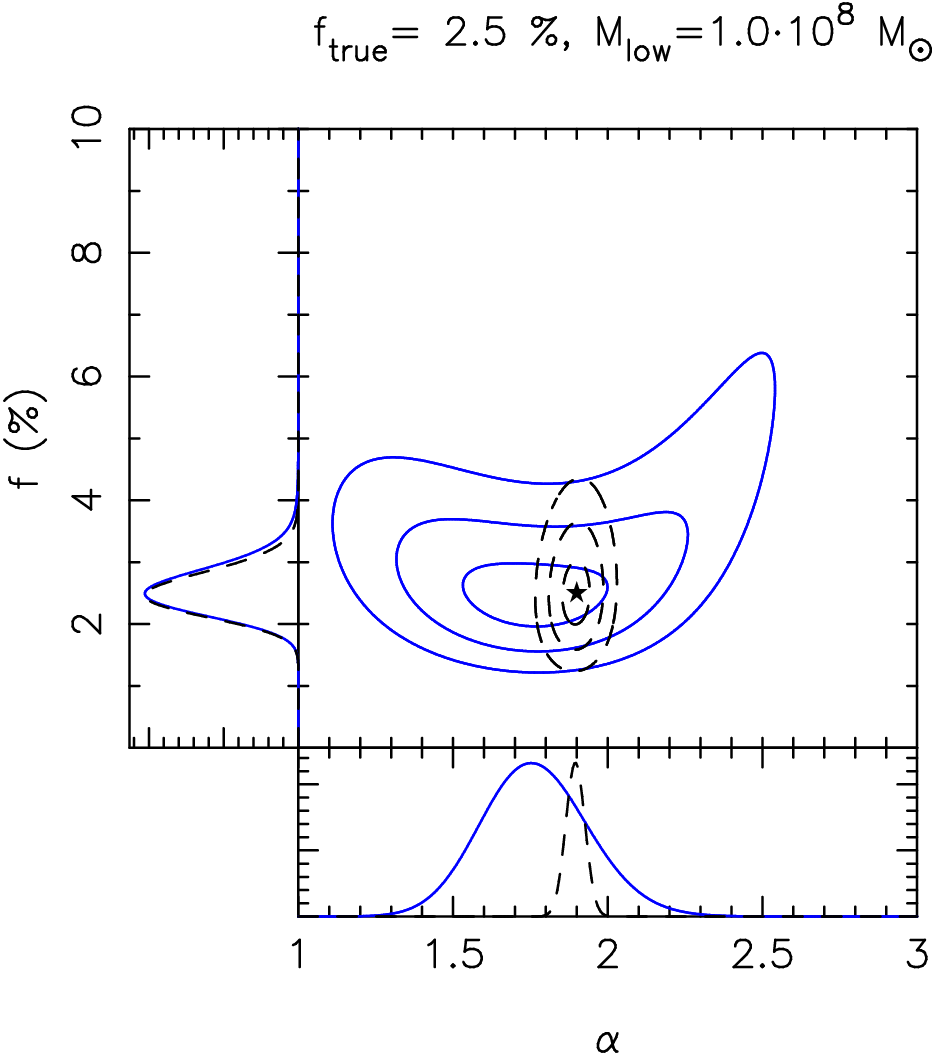}
\end{minipage}

\begin{minipage}[b]{0.33\linewidth}
\includegraphics[scale=0.25]{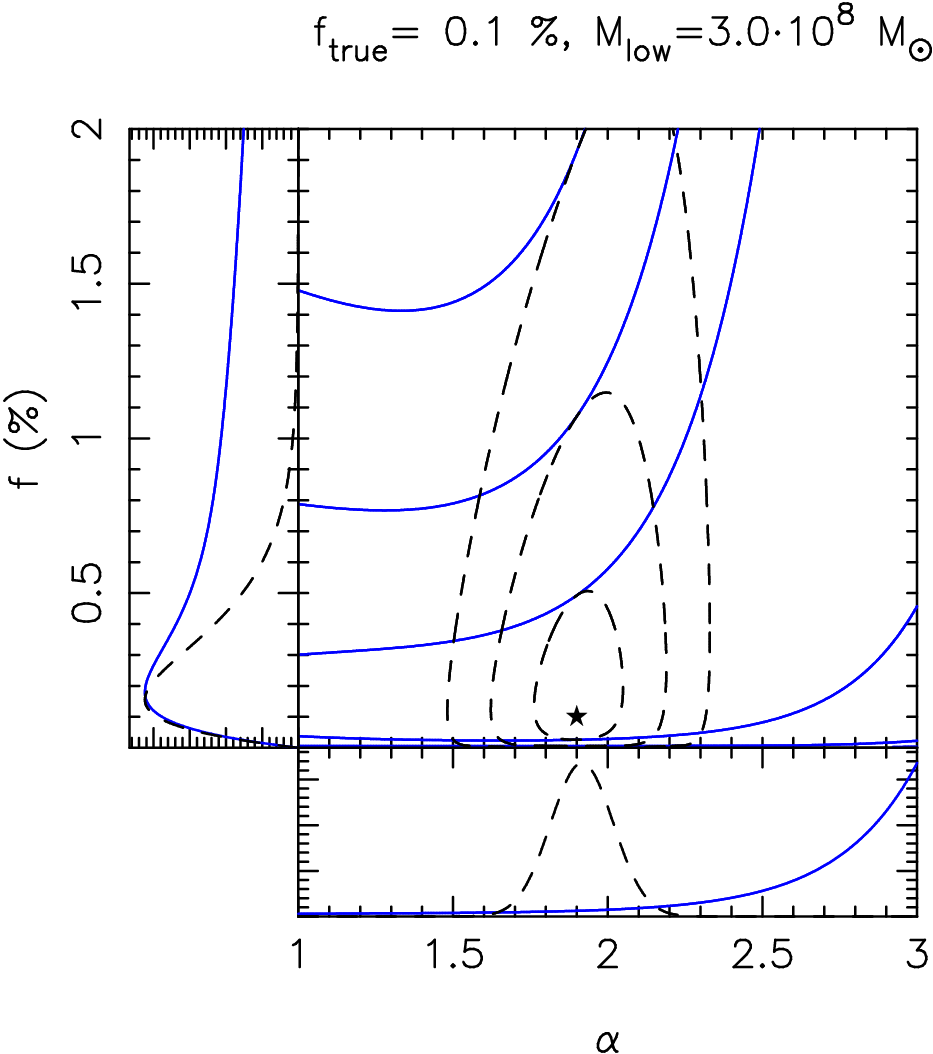}
\end{minipage}
\hspace{0.1cm}
\begin{minipage}[b]{0.33\linewidth}
\includegraphics[scale=0.25]{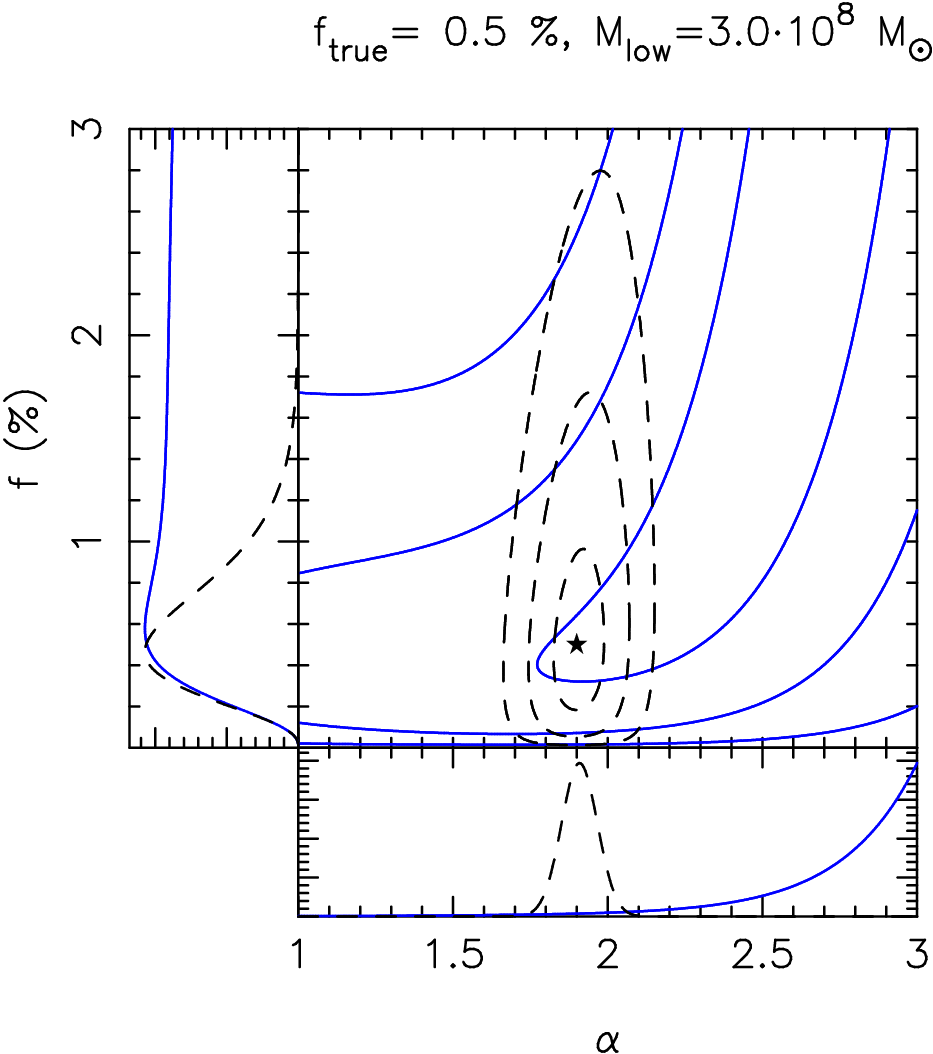}
\end{minipage}
\hspace{0.01cm}
\begin{minipage}[b]{0.25\linewidth}
\includegraphics[scale=0.25]{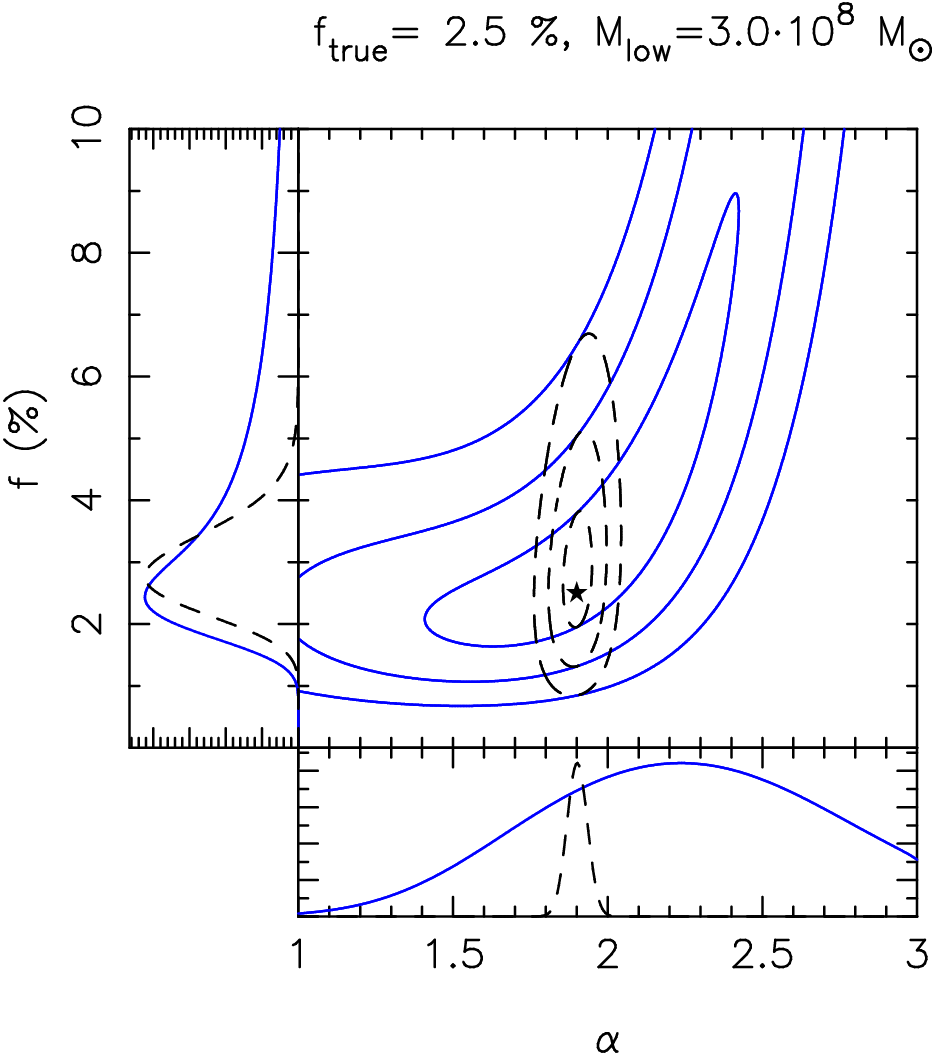}
\end{minipage}
\caption{Results for systems with 10 randomly realised lenses. In each panel, the joint probability $P\left( \alpha,f ~|~ \{n_s,\vect{m}\},\vect{p} \right)$ contours and marginalized probabilities 
$P\left( f ~|~\{n_s,\vect{m}\},\vect{p} \right)$ and $P\left( \alpha ~|~\{n_s,\vect{m}\},\vect{p} \right)$ are given for a uniform prior (solid lines) and for a Gaussian prior on $\alpha$ (dashed lines). Moving from one panel to next the substructure fraction $f$ increases from left to right and the detection limit $M_{\rm{low}}$ increases from top to bottom.}
\label{fig:lens_10}
\end{center}
\end{figure*}

\begin{figure*}
\begin{center}
\begin{minipage}[b]{0.33\linewidth}
\includegraphics[scale=0.25]{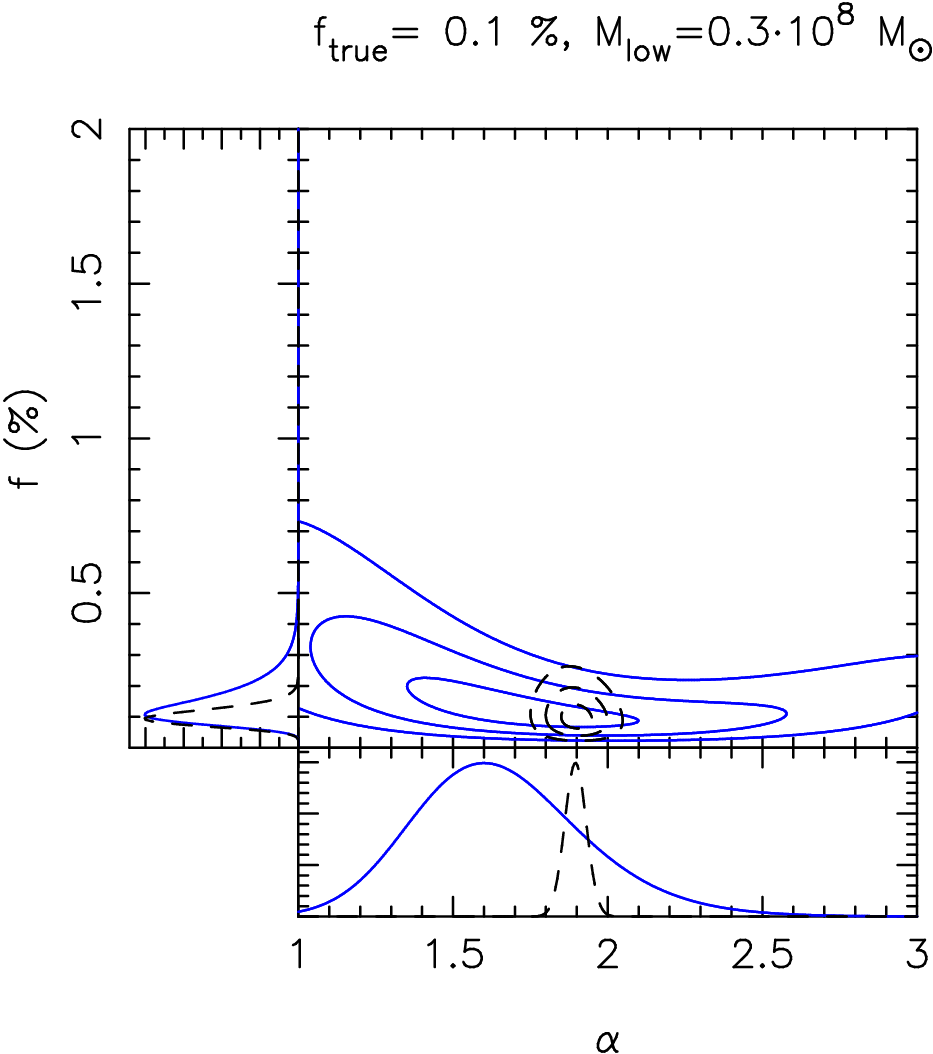}
\end{minipage}
\hspace{0.1cm}
\begin{minipage}[b]{0.33\linewidth}
\includegraphics[scale=0.25]{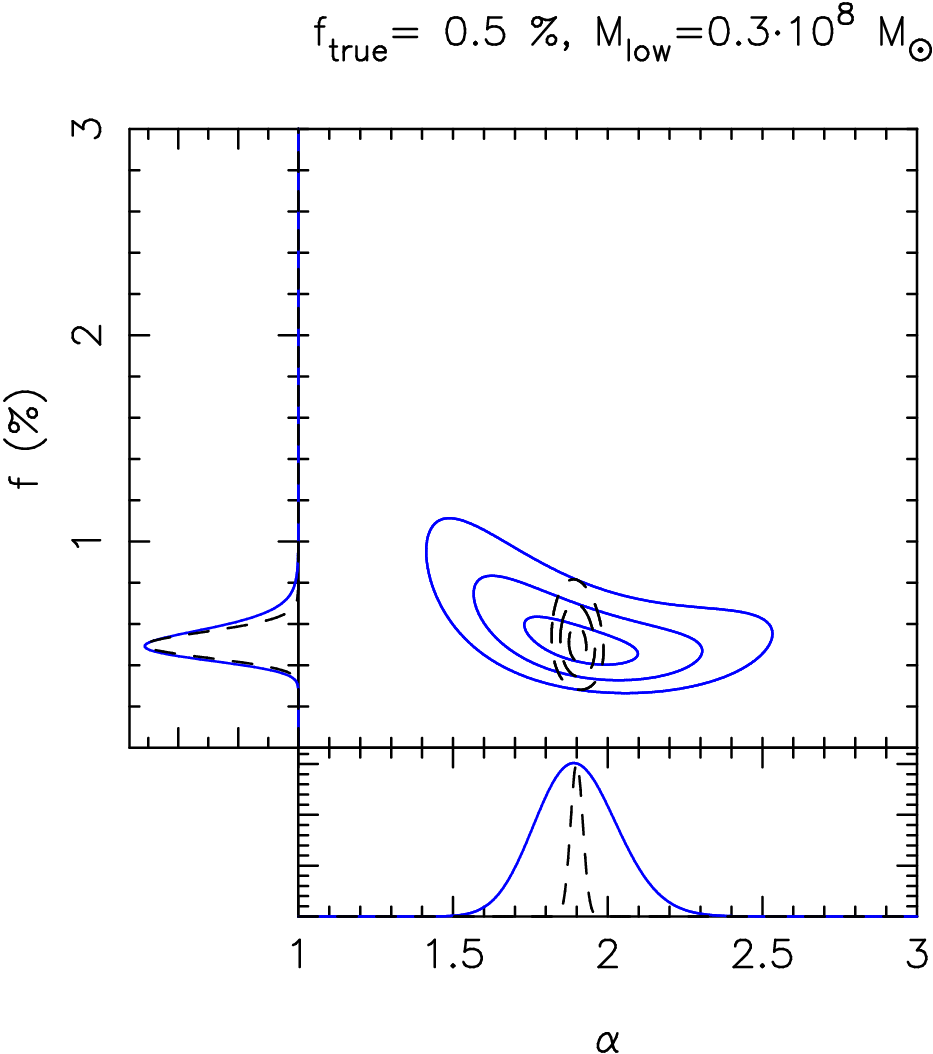}
\end{minipage}
\hspace{0.01cm}
\begin{minipage}[b]{0.25\linewidth}
\includegraphics[scale=0.25]{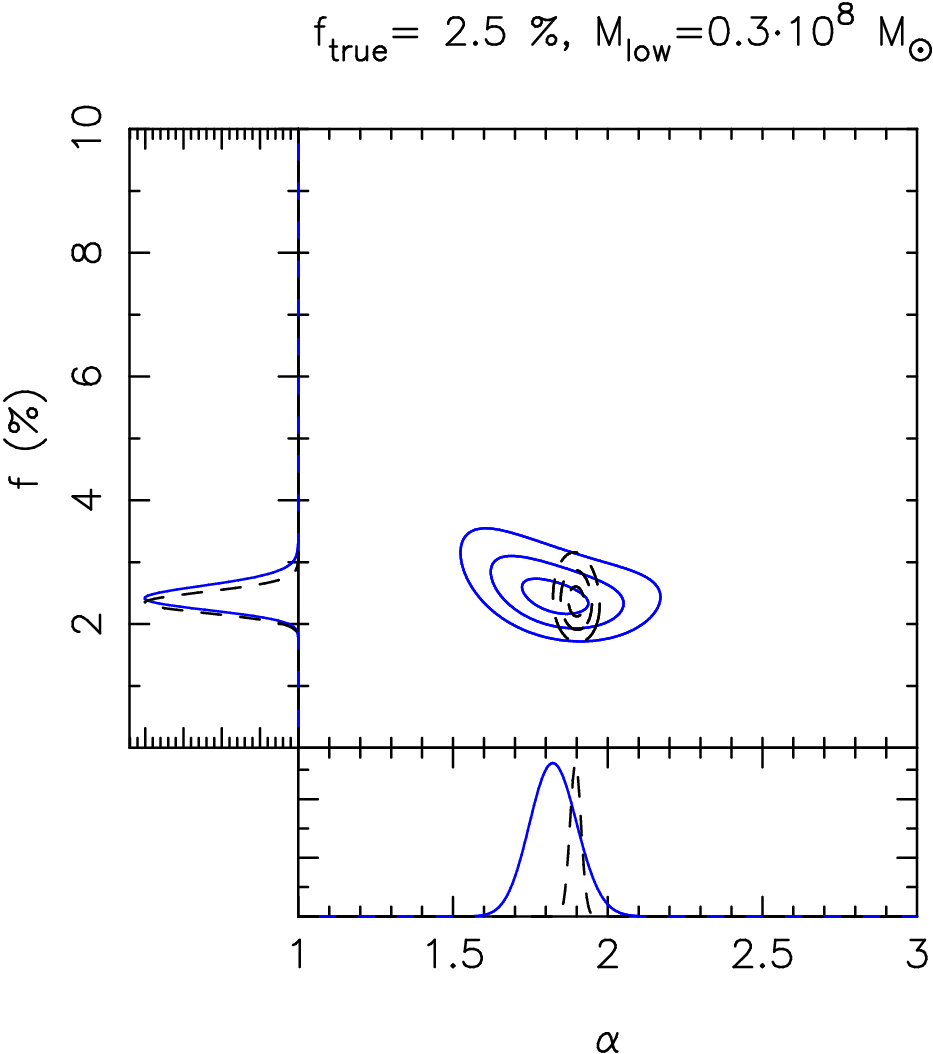}
\end{minipage}

\begin{minipage}[b]{0.33\linewidth}
\includegraphics[scale=0.25]{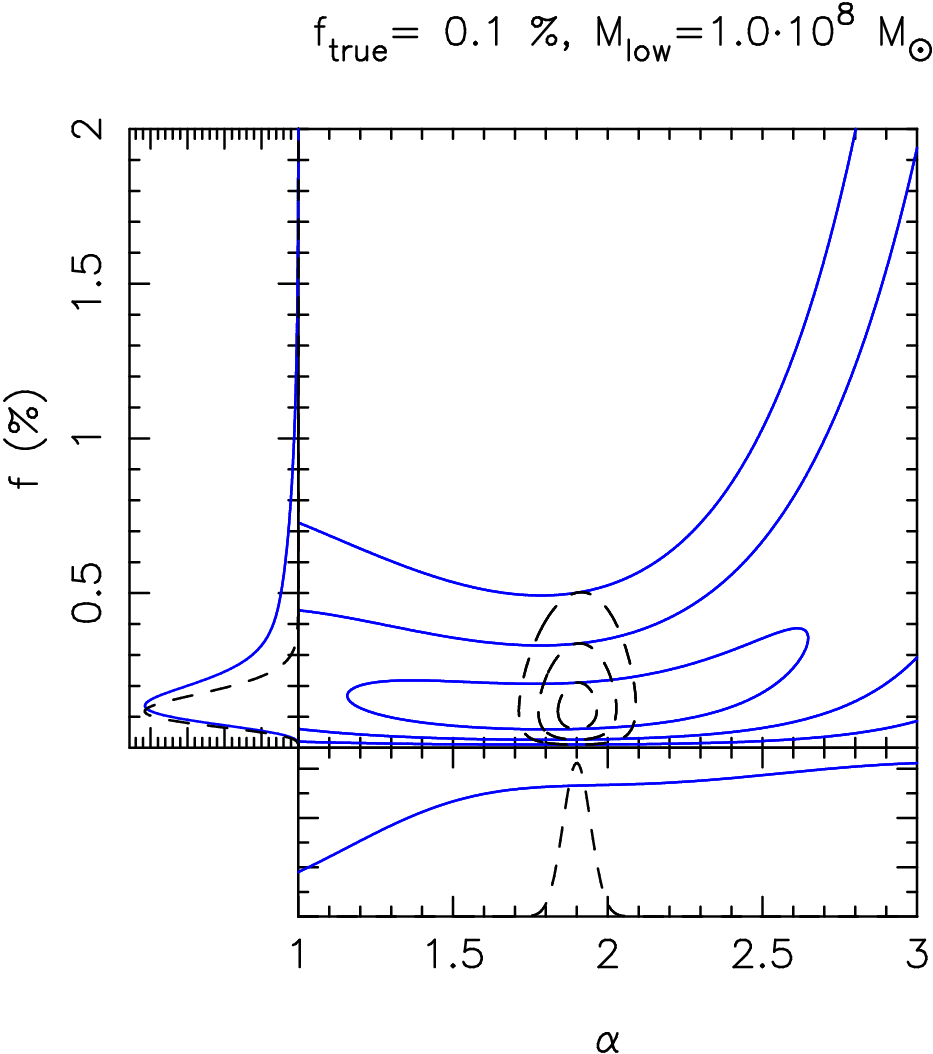}
\end{minipage}
\hspace{0.1cm}
\begin{minipage}[b]{0.33\linewidth}
\includegraphics[scale=0.25]{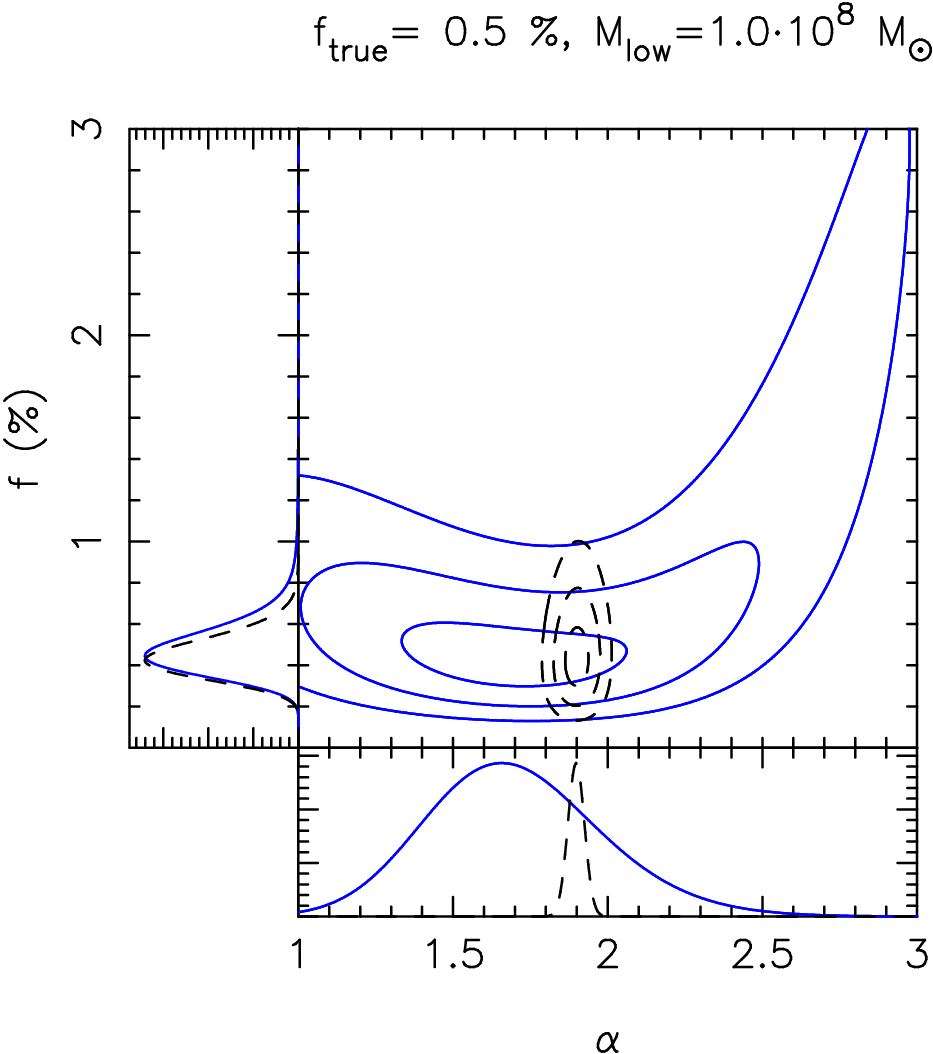}
\end{minipage}
\hspace{0.01cm}
\begin{minipage}[b]{0.25\linewidth}
\includegraphics[scale=0.25]{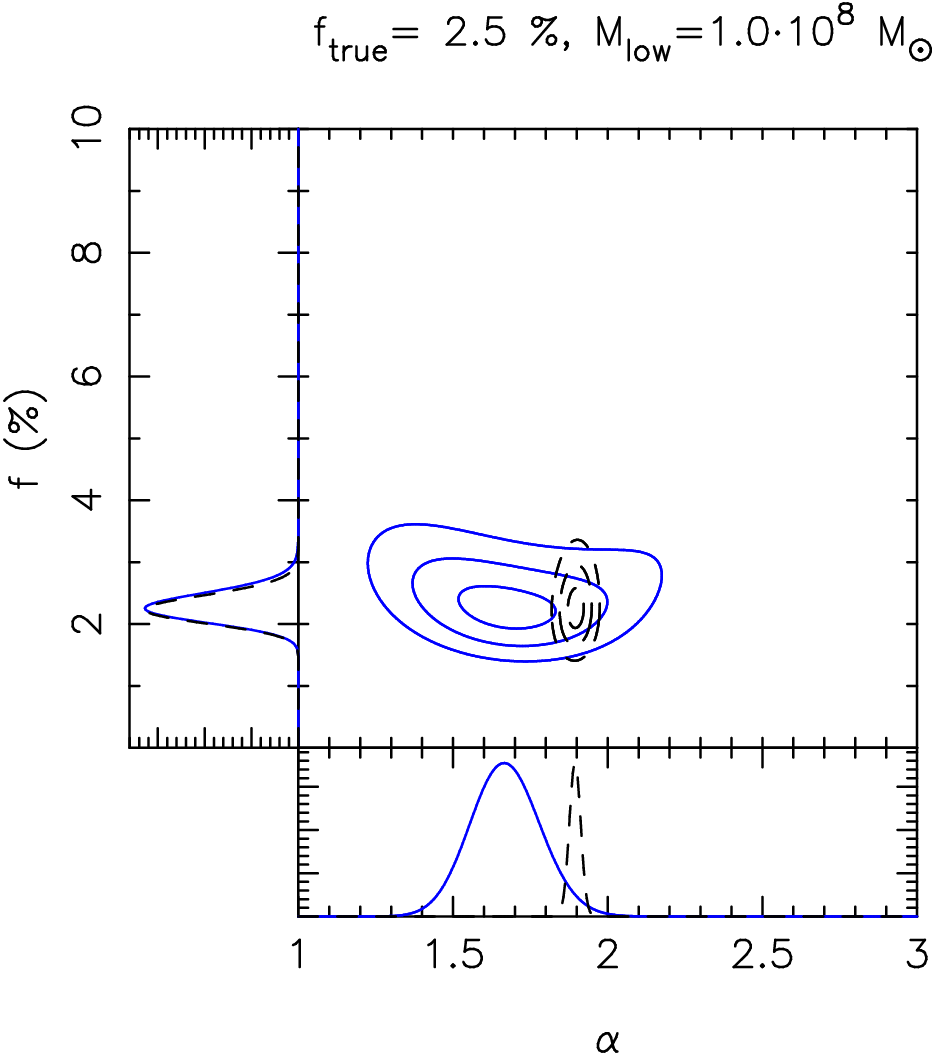}
\end{minipage}

\begin{minipage}[b]{0.33\linewidth}
\includegraphics[scale=0.25]{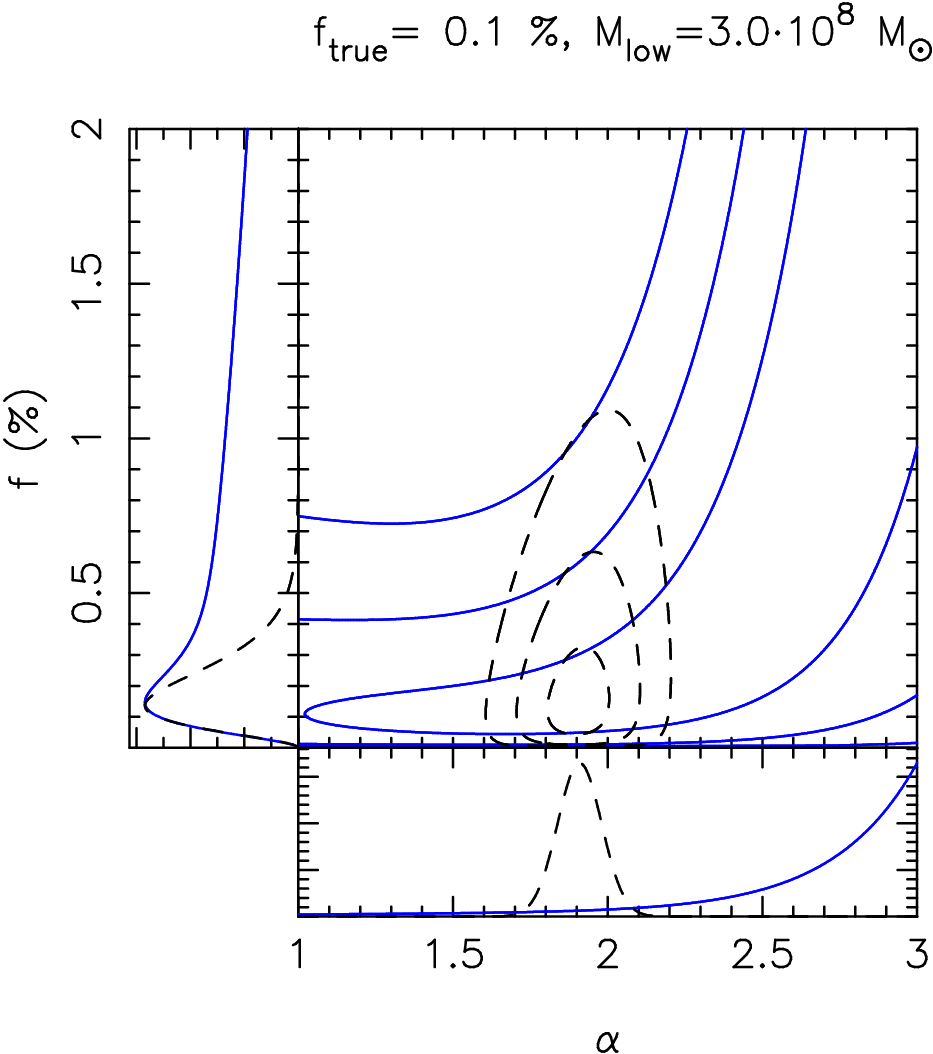}
\end{minipage}
\hspace{0.1cm}
\begin{minipage}[b]{0.33\linewidth}
\includegraphics[scale=0.25]{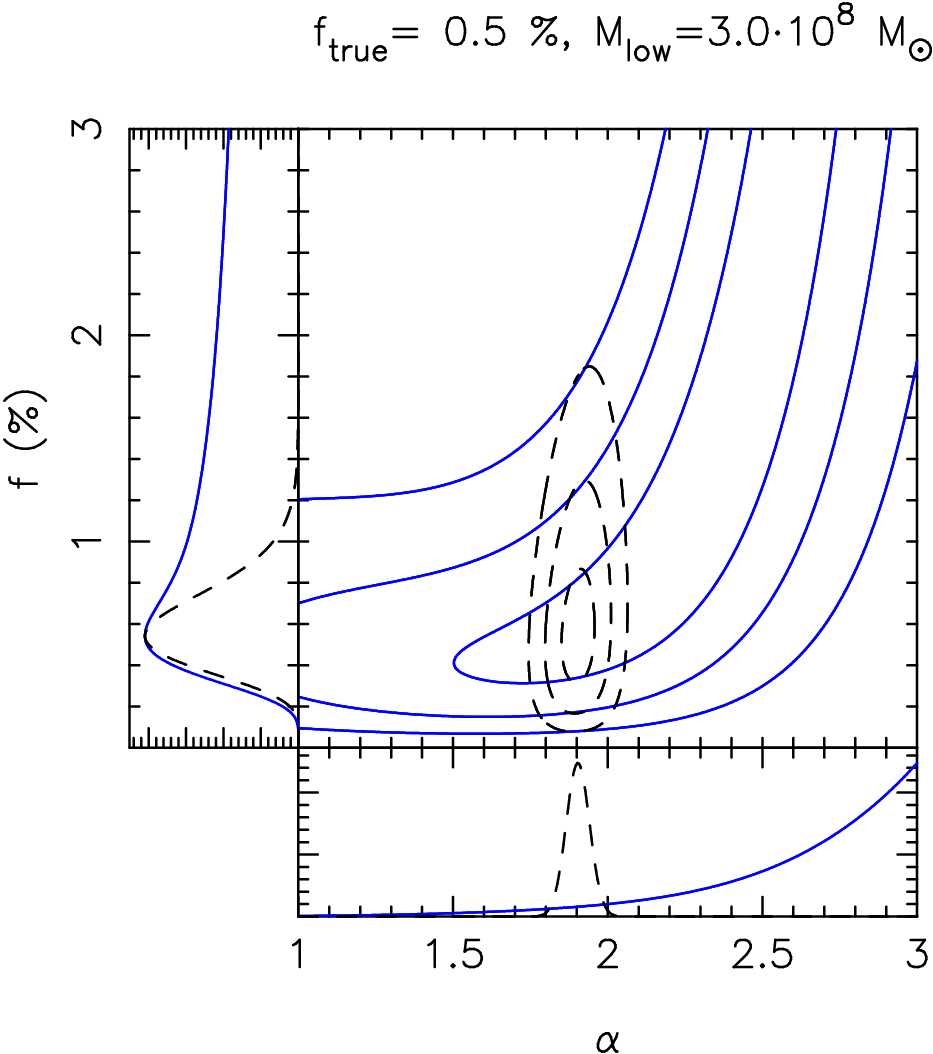}
\end{minipage}
\hspace{0.01cm}
\begin{minipage}[b]{0.25\linewidth}
\includegraphics[scale=0.25]{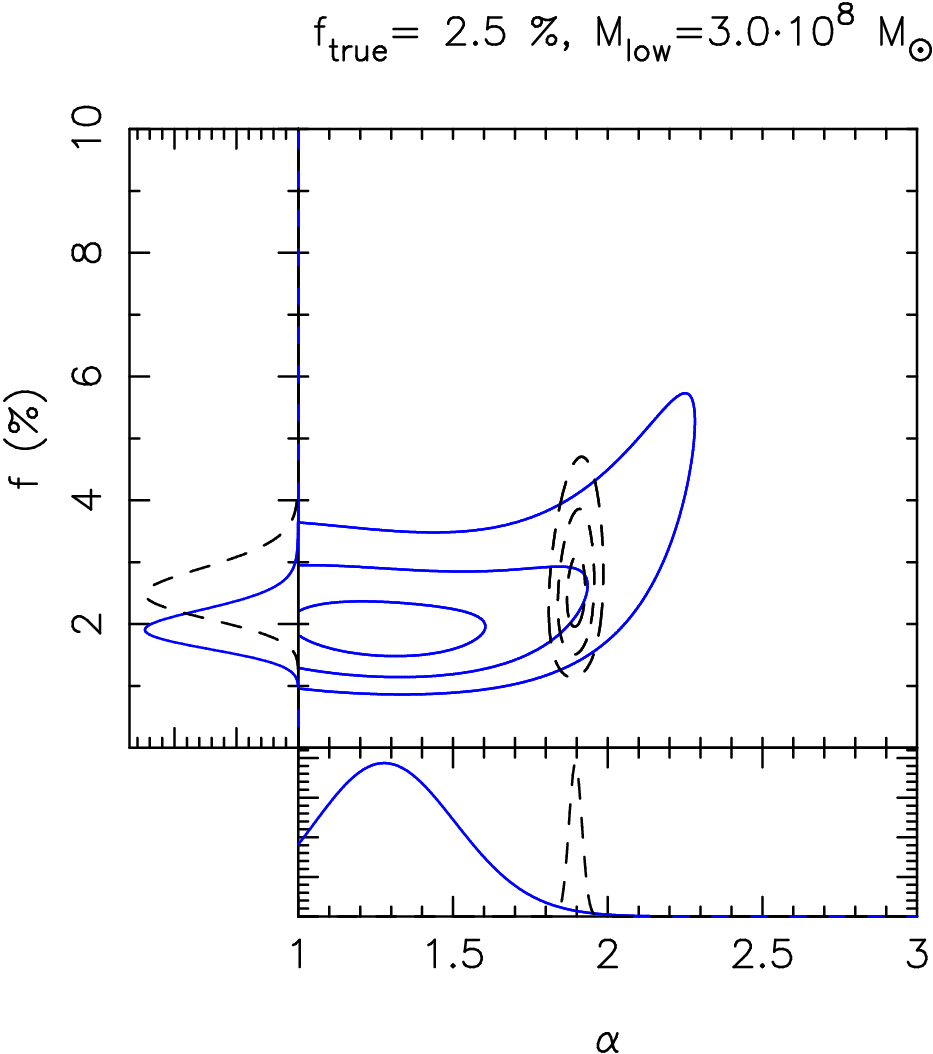}
\end{minipage}
\caption{Similar to Fig. \ref{fig:lens_10} for systems with 30 lenses.}
\label{fig:lens_30}
\end{center}
\end{figure*}

\begin{figure*}
\begin{minipage}[h]{0.3\linewidth}
\includegraphics[scale=0.3]{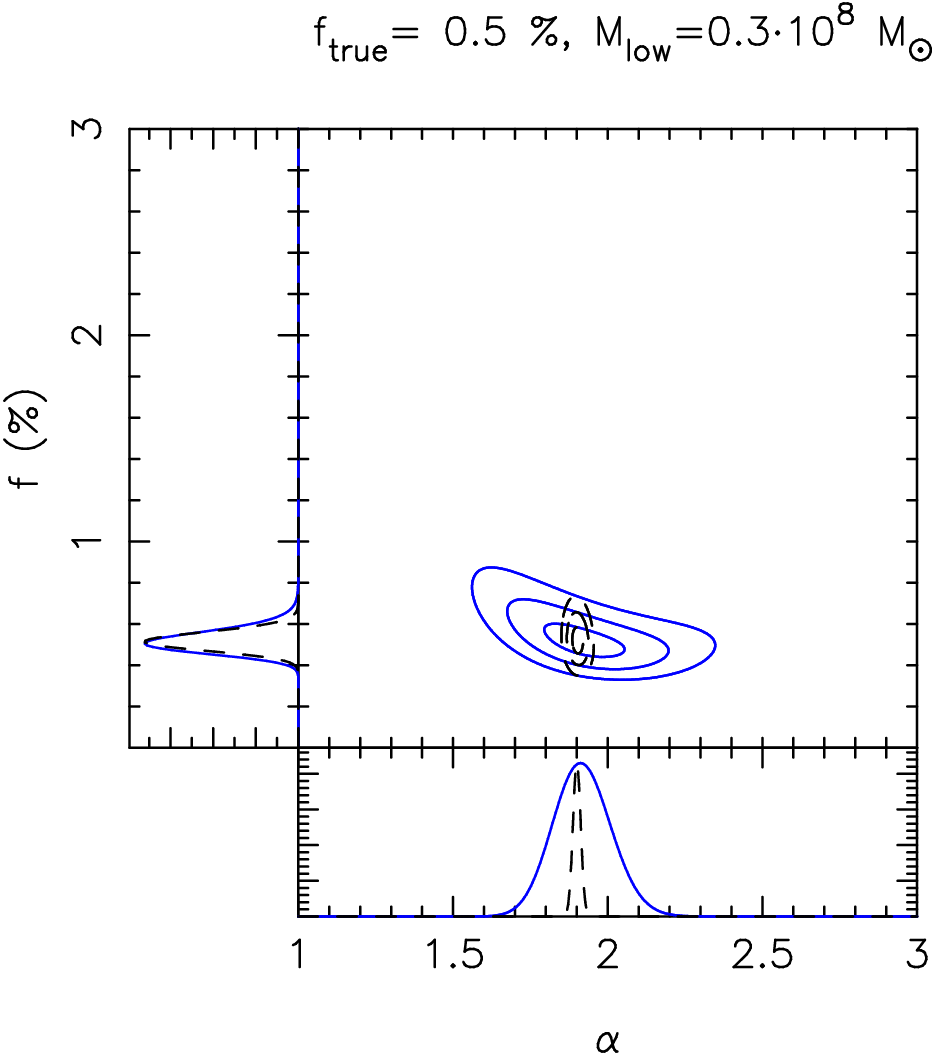}
\end{minipage}
\hspace{-0.55cm}
\begin{minipage}[h]{0.3\linewidth}
\includegraphics[scale=0.3]{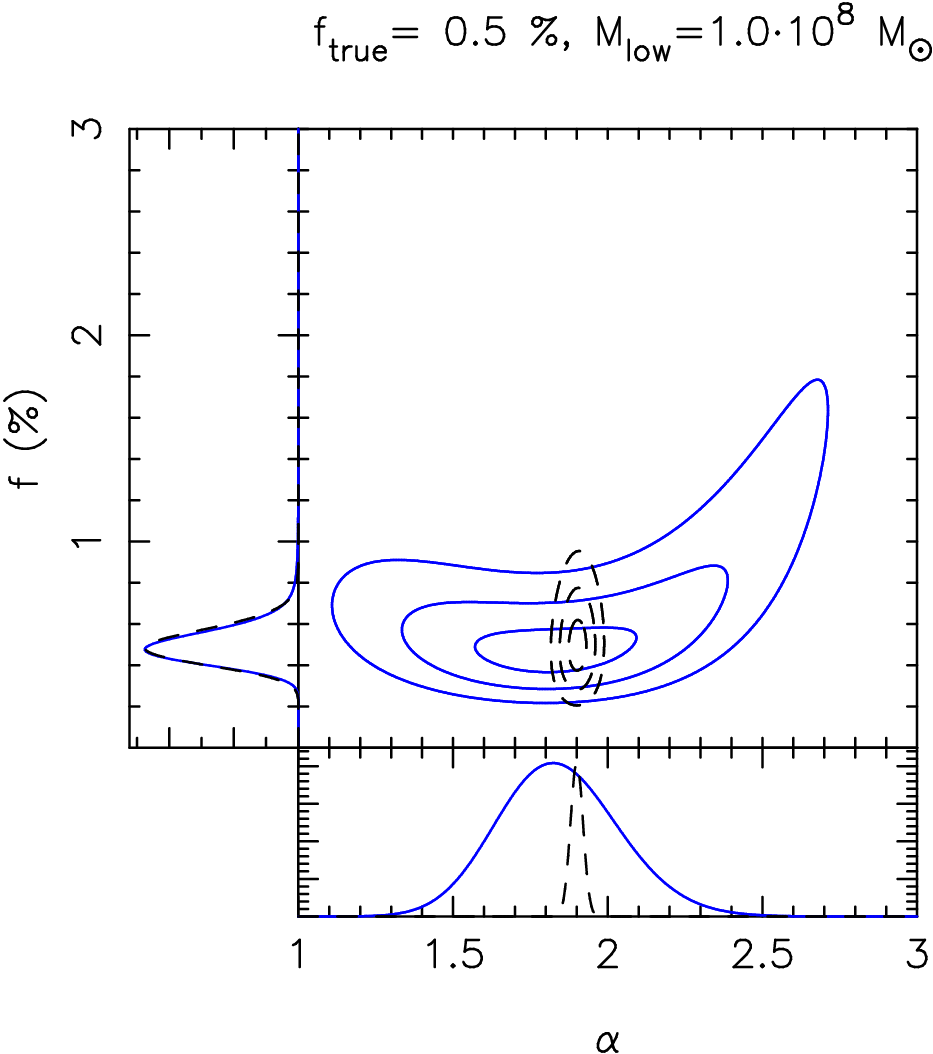}
\end{minipage}
\hspace{-0.56cm}
\begin{minipage}[h]{0.3\linewidth}
\includegraphics[scale=0.3]{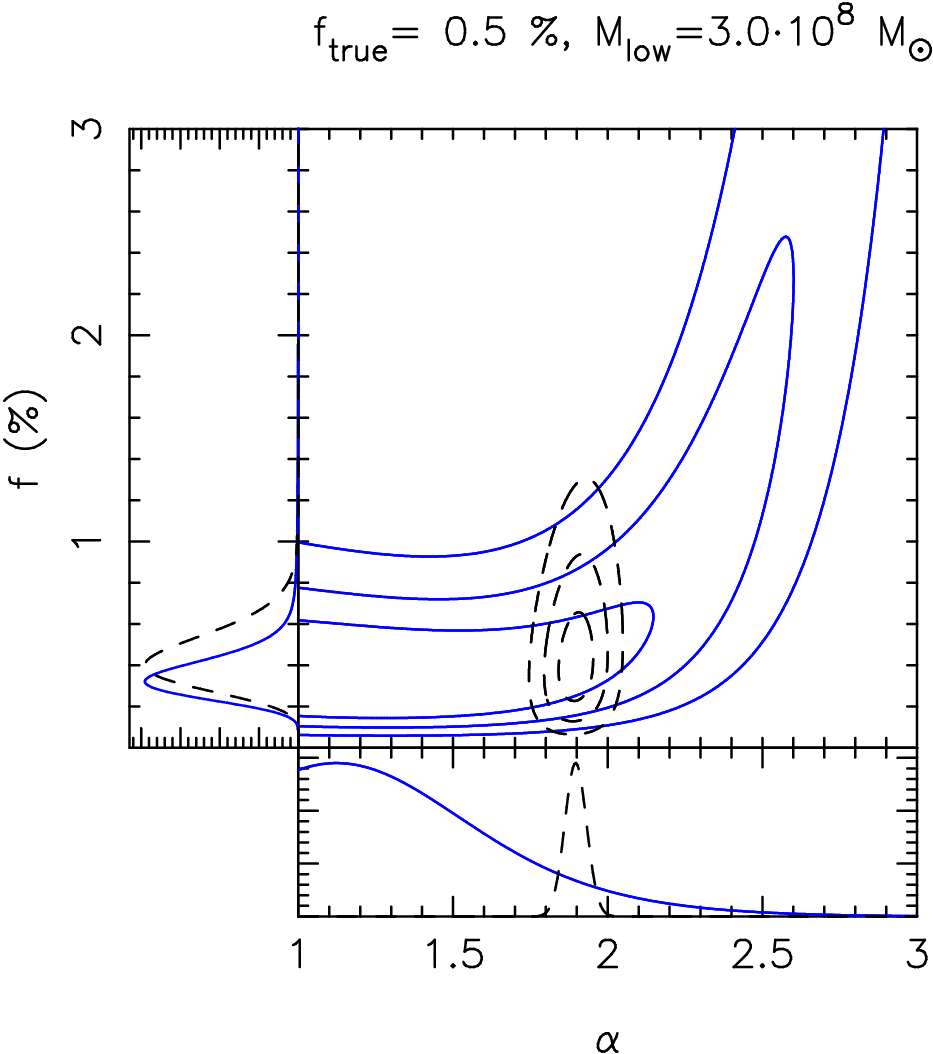}
\end{minipage}
\caption{Results for three samples with 200 randomly realised lenses with $f=0.5\%$ and respectively $M_{\rm{low}}=0.3\times 10^{8}\msun$ (left panel), $M_{\rm{low}}=1.0\times 10^{8}\msun$ (middle panel) and $M_{\rm{low}}=3.0\times 10^{8}\msun$ (right panel). The joint probability $P\left( \alpha,f ~|~ \{n_s,\vect{m}\},\vect{p} \right)$ contours and marginalized probabilities $P\left( f ~|~\{n_s,\vect{m}\},\vect{p} \right)$ and $P\left( \alpha ~|~\{n_s,\vect{m}\},\vect{p} \right)$ for a uniform prior (solid lines) and for a Gaussian prior in $\alpha$ (dashed lines) are shown.}
\label{fig:lens_200}
\end{figure*}

We present results for cases in which the errors on the mass measurements can be neglected. Specifically, this translates into convolving the mass distribution not with a Gaussian but with a delta function around $m_i$ in the equations of Section 2.\\
\noi In Fig. \ref{fig:lens_10} we show the joint probability contours $P\left( \alpha,f ~|~ \{n_s,\vect{m}\},\vect{p} \right)$ and the marginalized probabilities $P\left( f ~|~\{n_s,\vect{m}\},\vect{p} \right)$ and $P\left( \alpha ~| ~\{n_s,\vect{m}\},\vect{p} \right)$, for systems containing 10 randomly realised lenses. Specifically the plotted contours contain, in the limit of a Gaussian distribution, respectively, 68\%, 95\% and 97.2\% of the marginalized probability function.\\ 
\noi In the case of a uniform prior, while a good upper limit to $f$ can always be set, little can be said about the slope, which can only be constrained for a limited number of favourable physical and observational conditions, such as $(f=0.5\%,M_{\rm{low}}=0.3\times10^{8}\msun)$, $(f=2.5\%,M_{\rm{low}}=0.3\times10^{8}\msun)$ and $(f=2.5\%,M_{\rm{low}}=1.0\times10^{8}\msun)$. In Fig. \ref{fig:lens_30} an equivalent plot is presented for systems with $n_l=30$; although even more stringent limits can be given for $f$, we are still unable to recover the underlying mass function for most of the possible scenarios.\\
\noi The situation can be substantially improved by increasing the number of detectable substructures with a larger number of lenses. To provide insight into future capabilities, results from three samples of 200 lenses with $f=0.5\%$ and respectively $M_{\rm{low}}=0.3\times 10^{8}\msun$, $M_{\rm{low}}=1.0\times 10^{8}\msun$, and $M_{\rm{low}}=3.0\times 10^{8}\msun$ are given in Fig. \ref{fig:lens_200}. Currently, no uniform sample with 200 lenses with high signal-to-noise ratio and high resolution (equivalent to that of HST) is available. However, forthcoming ground and space-based instruments (e.g LSST/JDEM, EVLA, e-Merlin, LOFAR and SKA) can provide these numbers (in fact beyond these) in the coming 5--10 years and the required data quality by dedicated follow-up. A detailed characterisation of the CDM mass-function, through the technique of \citet{Vegetti09}, could therefore be realisable in the coming years if investments are made in large high-resolution and high-sensitivity lens surveys with these instruments \citep[see][]{Koopmans09}.
As can be seen from both Figs. \ref{fig:lens_10} and \ref{fig:lens_30}, in the case of a Gaussian prior on $\alpha$, tight limits can be set on the mass fraction for all possible combinations of the considered parameters.\\
\noi The results from this section are summarised in Table 1, where we report the input values for each parameter, the recovered maximum-posterior values $\left(f_{\rm{MP}}; \alpha_{\rm{MP}}\right)$ at which $P\left( \alpha,f ~|~ \{n_s,\vect{m}\},\vect{p} \right)$ reaches its maximum, the median, the 68\% and 95\% ($\sigma_{68}$ and $\sigma_{95}$), confidence levels of the marginalized probabilities $P\left( f ~|~\{n_s,\vect{m}\},\vect{p} \right)$ and $P\left( \alpha ~|~ \{n_s,\vect{m}\},\vect{p} \right)$ for both the cases of a uniform and a Gaussian prior on $\alpha$. $\sigma_{68}$, and $\sigma_{95}$, in the particular case of a Gaussian distribution, respectively represent the $1\sigma$ and $2\sigma$ error.

\subsection{The effect of mass measurement errors}
\begin{figure*}
\begin{center}
\begin{minipage}[b]{0.33\linewidth}
\includegraphics[scale=0.25]{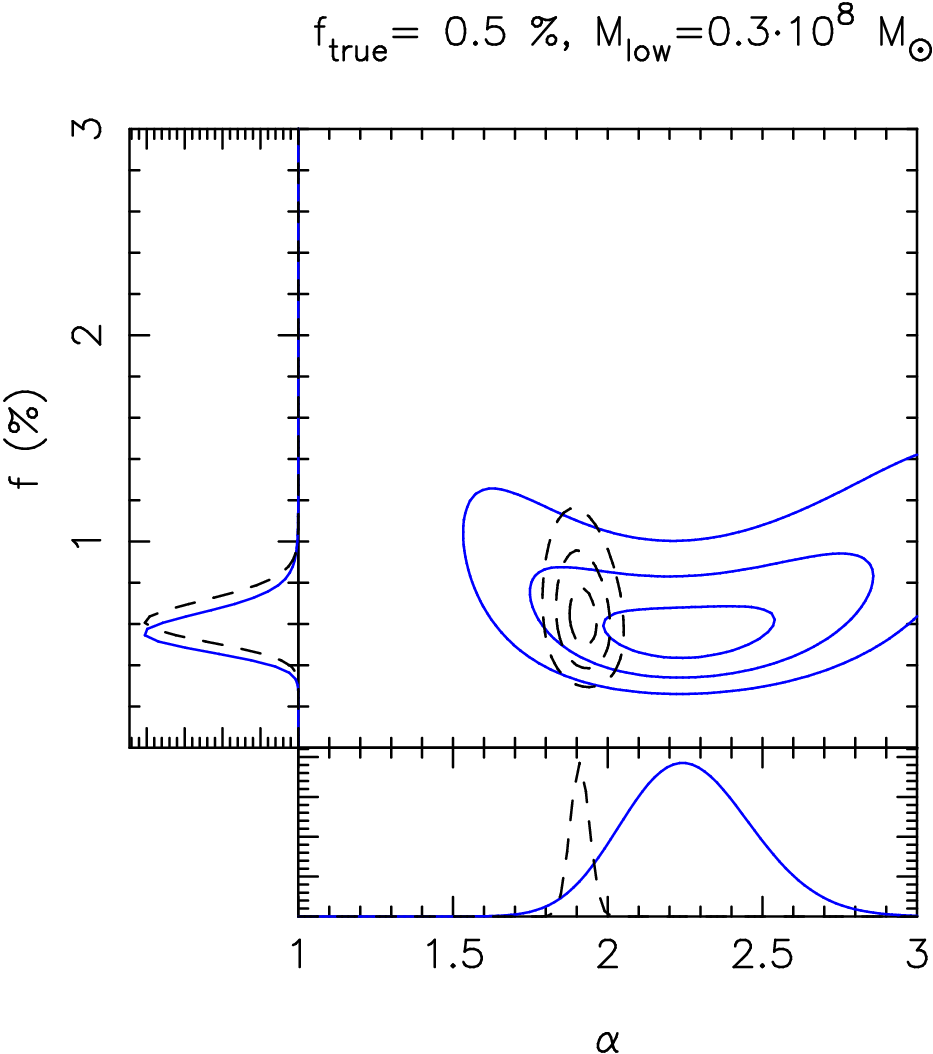}
\end{minipage}
\hspace{0.1cm}
\begin{minipage}[b]{0.33\linewidth}
\includegraphics[scale=0.25]{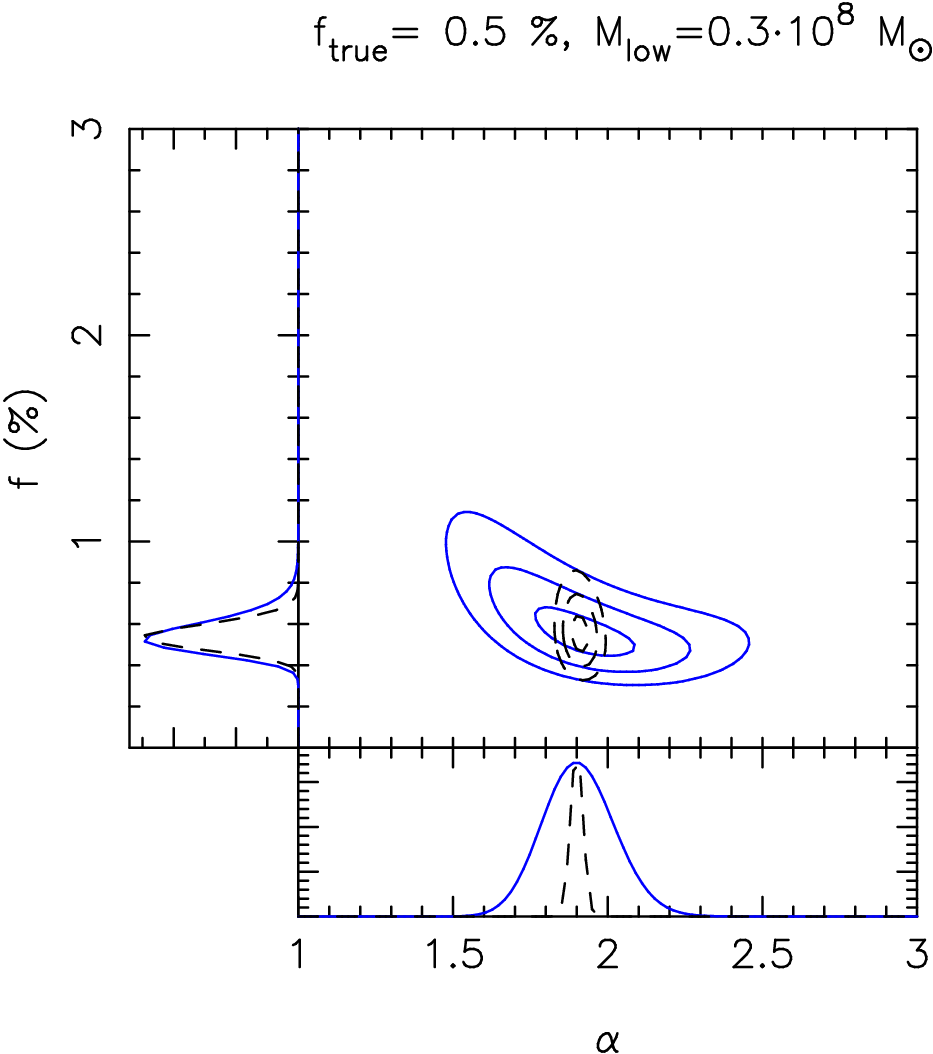}
\end{minipage}
\hspace{0.01cm}
\begin{minipage}[b]{0.25\linewidth}
\includegraphics[scale=0.25]{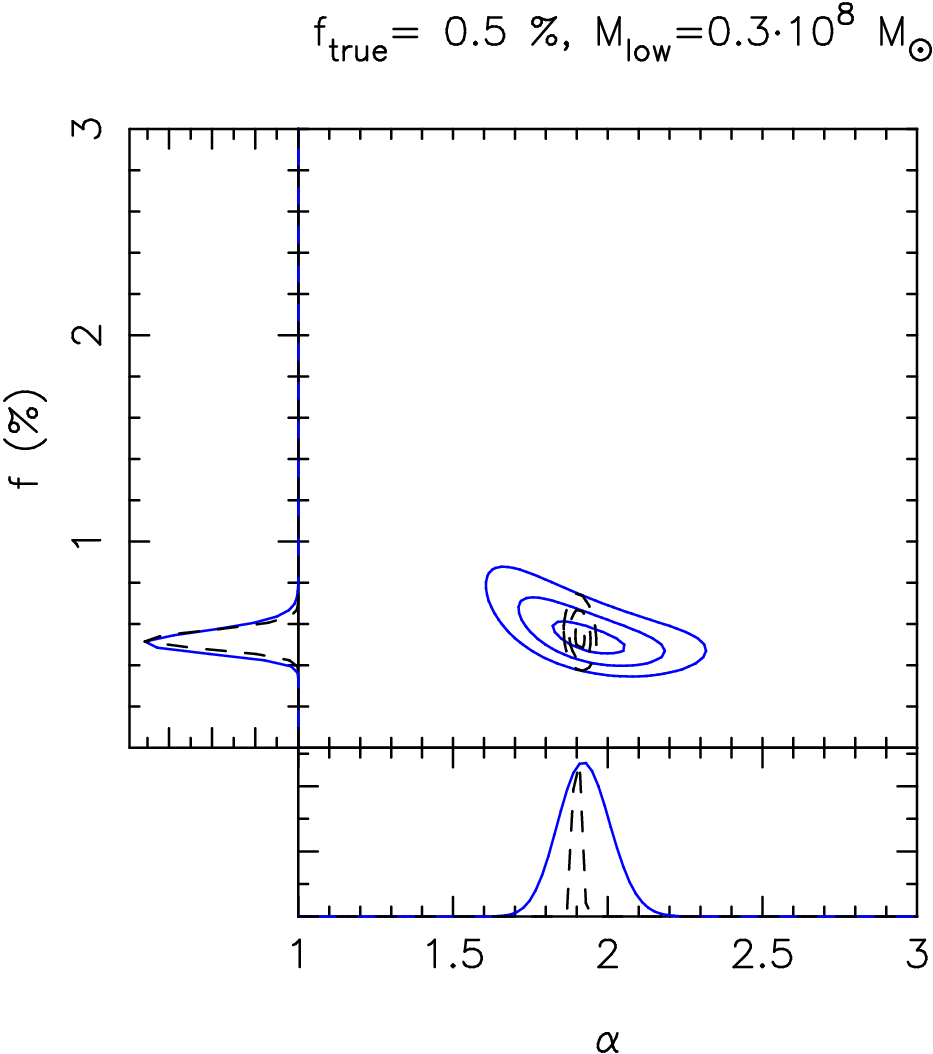}
\end{minipage}

\begin{minipage}[b]{0.33\linewidth}
\includegraphics[scale=0.25]{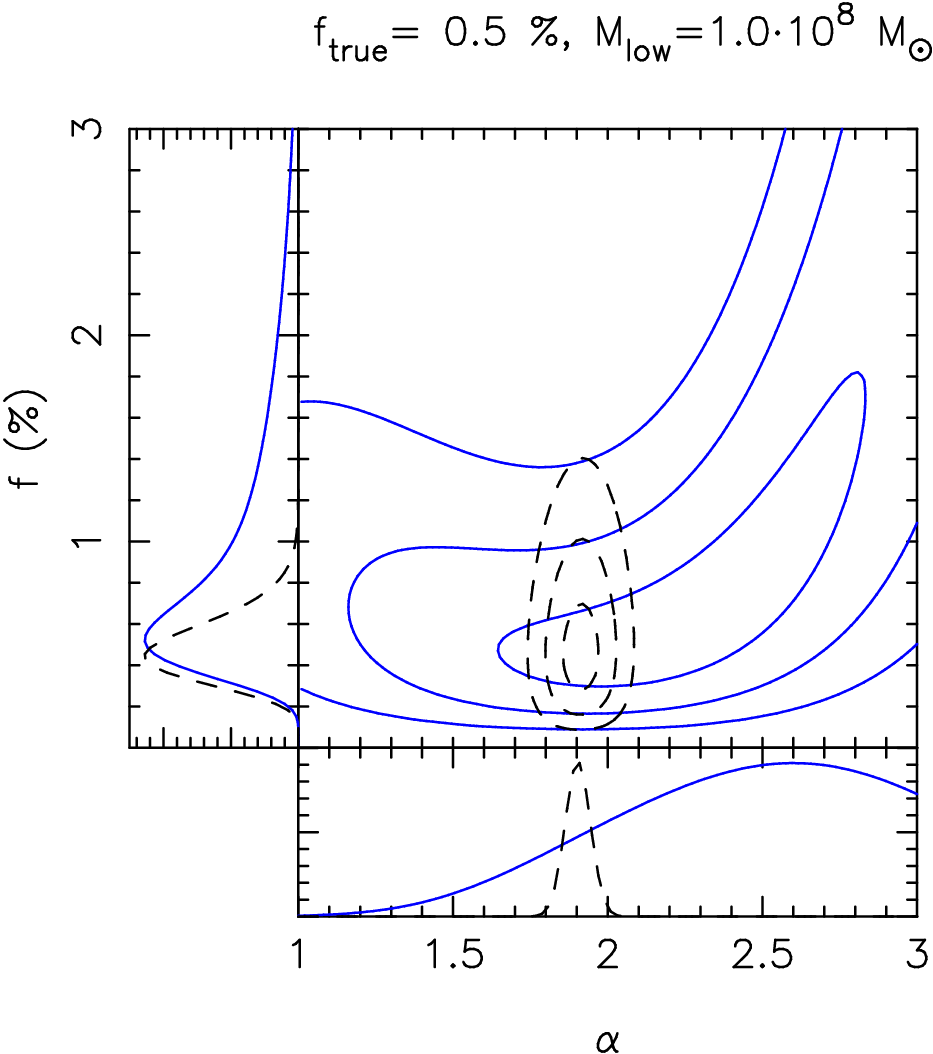}
\end{minipage}
\hspace{0.1cm}
\begin{minipage}[b]{0.33\linewidth}
\includegraphics[scale=0.25]{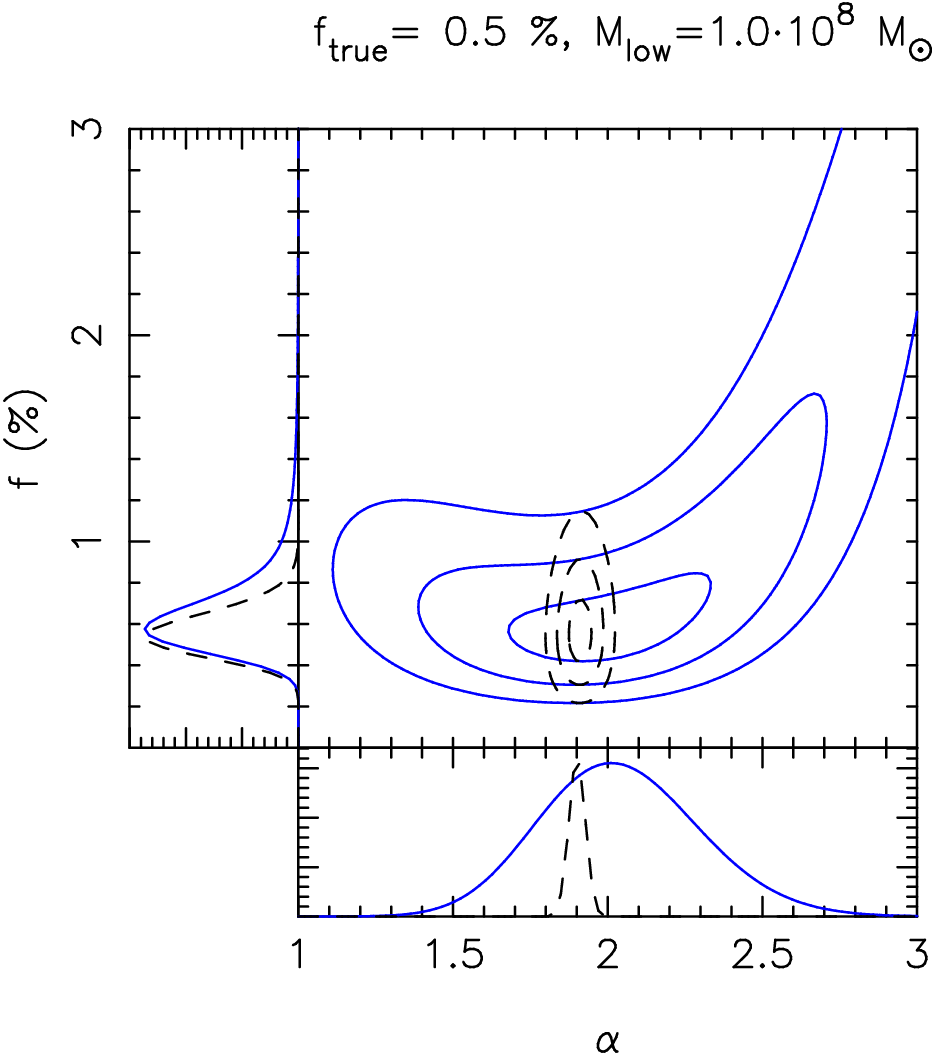}
\end{minipage}
\hspace{0.01cm}
\begin{minipage}[b]{0.25\linewidth}
\includegraphics[scale=0.25]{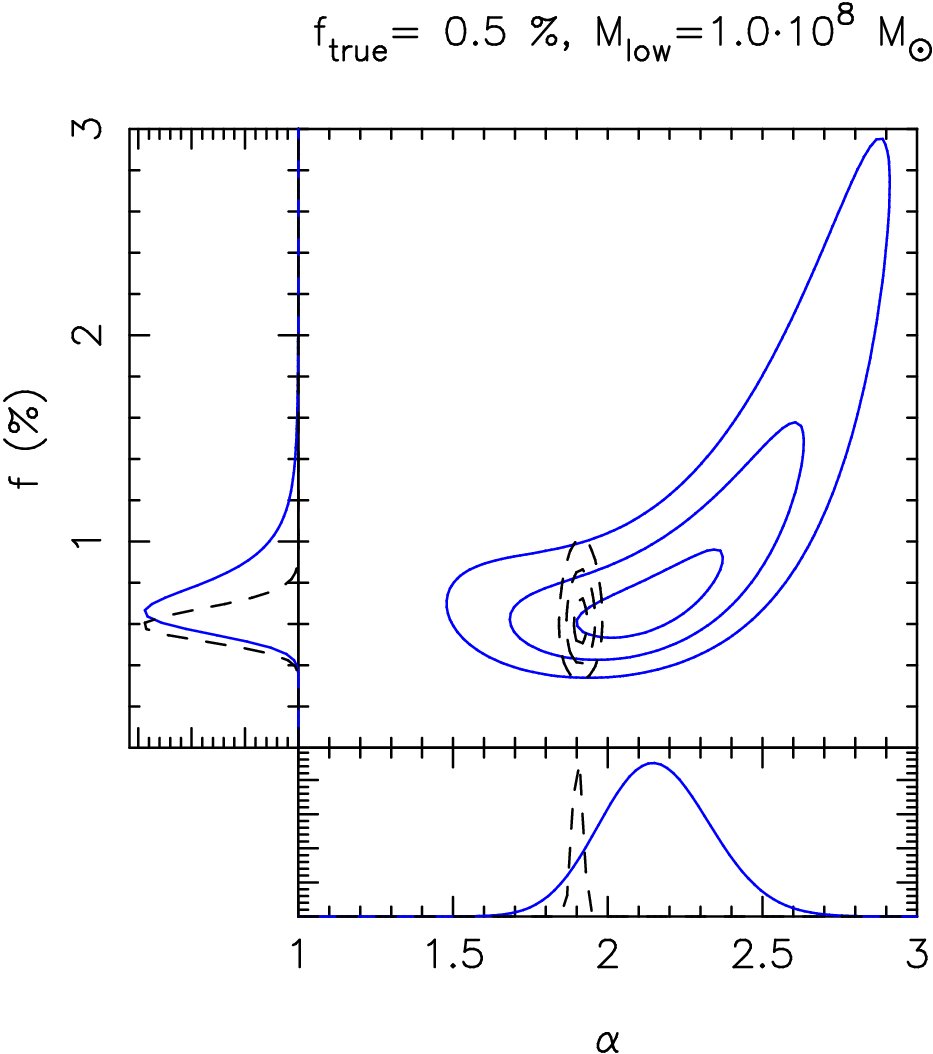}
\end{minipage}

\begin{minipage}[b]{0.33\linewidth}
\includegraphics[scale=0.25]{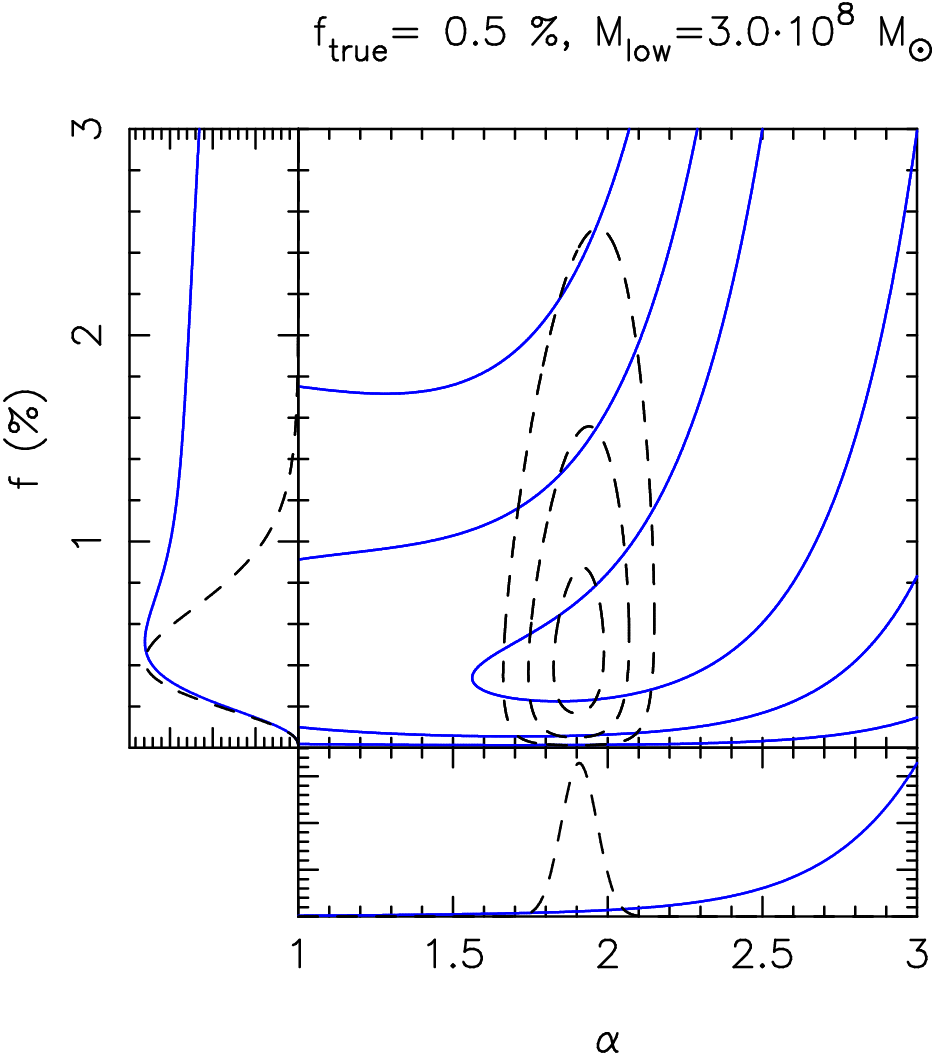}
\end{minipage}
\hspace{0.1cm}
\begin{minipage}[b]{0.33\linewidth}
\includegraphics[scale=0.25]{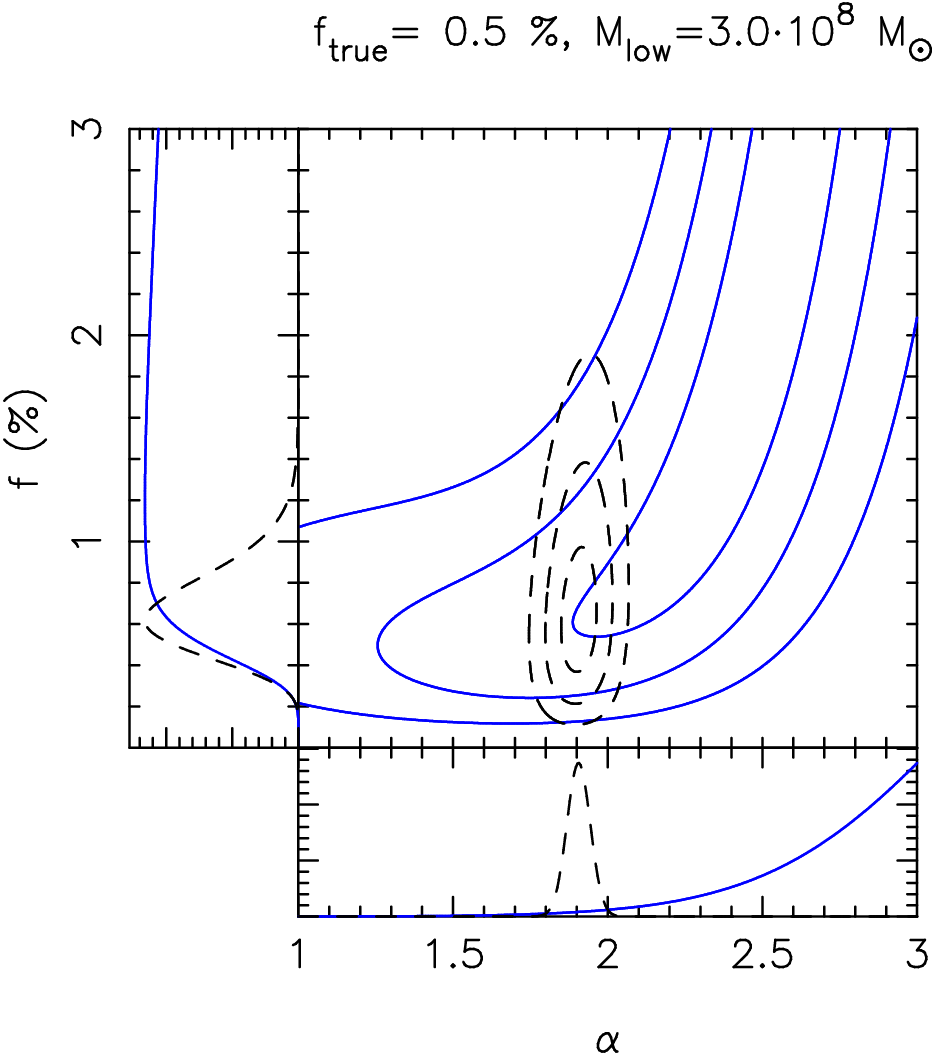}
\end{minipage}
\hspace{0.01cm}
\begin{minipage}[b]{0.25\linewidth}
\includegraphics[scale=0.25]{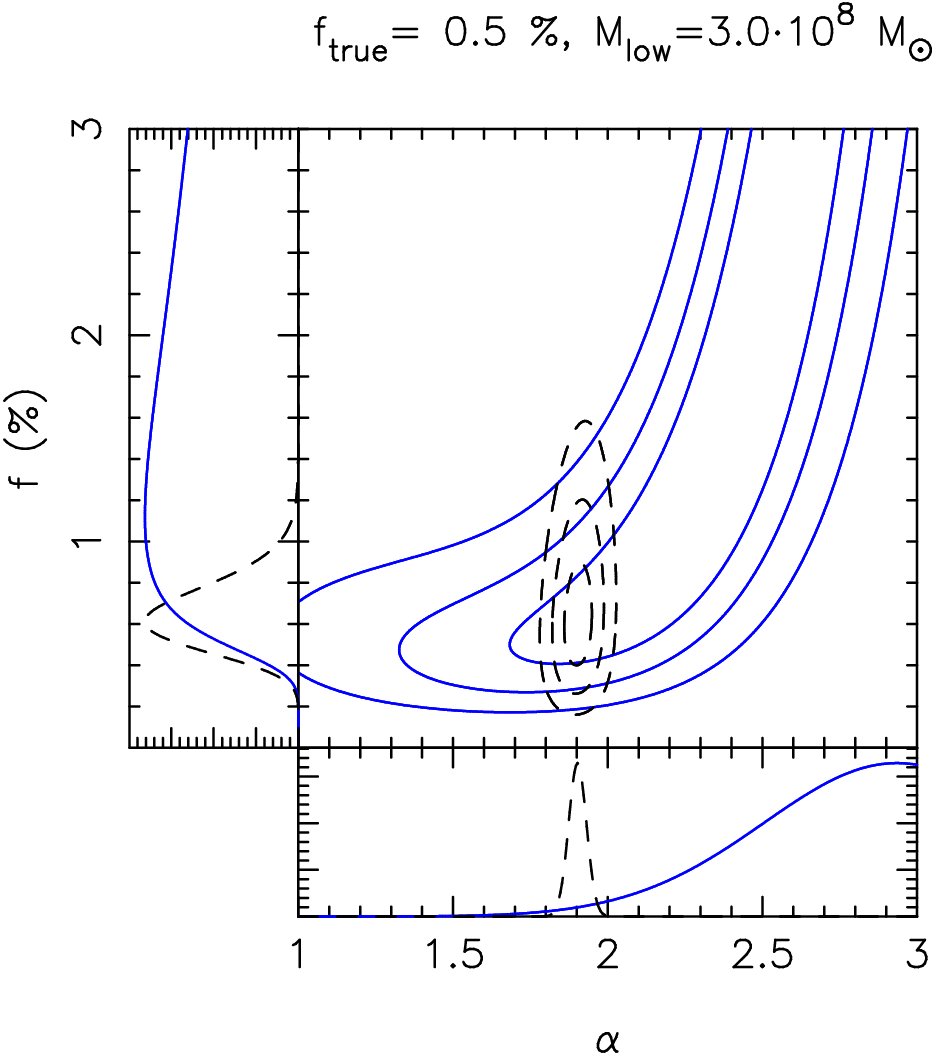}
\end{minipage}

\caption{Effect of different measurement error levels on the substructure mass. Similar as Fig. \ref{fig:lens_10} for systems with $f=0.5\%$, $M_{\rm{low}}=0.3\times 10^{8}\msun$ and $\sigma_{m}=0.1\times10^{8}\msun$ (upper panels), with $M_{\rm{low}}= 10^{8}\msun$ and $\sigma_{m}=1/3\times 10^{8}\msun$ (middle panels) and with $M_{\rm{low}}= 3\times 10^{8}\msun$ and $\sigma_{m}=1.0\times 10^{8}\msun$ (lower panels). Results for 10, 30 and 200 lenses are plotted in the left, middle and right panels, respectively. }
\label{fig:mass_error}
\end{center}
\end{figure*}

We explore now how the presence of uncertainty in the mass measurements affects our analysis.
In particular, we consider three cases: $\sigma_{m}=0.1\times10^{8}\msun$ with $M_{\rm{low}}= 0.3\times10^{8}\msun$, $\sigma_{m}=1/3\times 10^{8}\msun$ with $M_{\rm{low}}= 10^{8}\msun$, and $\sigma_{m}=1.0\times10^{8}\msun$ with $M_{\rm{low}}= 3.0\times10^{8}\msun$ i.e. a limiting signal-to-noise ratio of $M_{\rm{low}}/\sigma_m= 3$. The lens systems analysed here have $n_l=10,30,200$ lenses and a mass fraction in substructures $f_{\rm{true}}=0.5\%$. Relative likelihood contours are plotted in the three panels of Fig. \ref{fig:mass_error}. These have to be compared with the equivalent \emph{no-error} results in Figs. \ref{fig:lens_10}, \ref{fig:lens_30} and \ref{fig:lens_200}. Results are reported in Table 2.\\
\noi The effect of measurement errors on the substructure mass depends on the form of prior adopted for $\alpha$, with uniform priors being more strongly affected than Gaussian ones. Errors as large as $\sigma_{m}=1/3\times 10^{8}\msun$ combined with mass threshold of $M_{\rm{low}}= 10^{8}\msun$, can significantly influence even systems with 200 lenses . 
\noi Specifically, these systems were created by drawing masses between $M_{\rm{min}}$ and $M_{\rm{max}}$, then scattering each mass with a Gaussian distribution (i.e. mimicking the measurement uncertainty) and then using only those objects that fall within the detection range $\left[M_{\rm{low}}, M_{\rm{high}}\right]$ to constrain the fraction and the mass distribution. We refer to the Appendix for a mathematical proof that this way of proceeding is equivalent to drawing between $M_{\rm{low}}$ and $M_{\rm{high}}$ with a Poisson probability density distribution of expectation values given by equation (\ref{equ:mu}).

\section{Conclusions}
We have introduced a statistical formalism for the interpretation and the generalisation of subhalo detection in gravitational lens galaxies, that allows us to quantify the mass fraction and the mass function of CDM substructures. Given mock sets of lenses, with properties typical of a CDM cosmology, we have analysed how well the true parameters can be recovered. The formalism depends on several parameters, such as e.g. the number of lenses, the mass detection threshold and the measurement errors. It has a very general nature and, in principle, it could be used to statistically analyse substructure detection by flux ratio anomalies or timedelay/astrometric perturbations as well. In practice, these methods would first need to show that their mass estimates are meaningful and second they would have to determine the probability distribution of flux-ratio anomalies or perturbations, either as function of the lens geometry or marginalized over all model parameters, which could be rather computationally expensive. The method has been tested on several mock data sets, with parameter settings based on our knowledge of the SLACS lenses. Several physical and observational scenarios have been considered. We list here the main results:
\begin{itemize}
\item If the number of arc/ring lens systems is $\ll 100$, as is the case for current surveys (e.g. SLACS), the ability to constrain the mass fraction and the mass function of satellites still depends on the form of prior which is assumed for $\alpha$. In particular, if results from numerical simulations are assumed to hold and a Gaussian prior with $\alpha_{\rm{true}}=1.9\pm 0.1$ is adopted, we are able to constrain both $\alpha$ and $f$ for any data sets containing a number of lenses $n_l\ge 10$, with improved limits for either increasing mass fractions, decreasing detection threshold or increasing number of lenses. If instead a wider range of possibilities is explored by assuming a uniform prior, one can still set strong limits on $f$, even for values as low as $f=0.1\%$ and a detection threshold $M_{\rm{low}}=0.3\times 10^{8}\msun$, but the mass function slope can be recovered only in a limited number of favourable cases, characterized by high mass fraction and low detection threshold.
 
\item Our ability to constrain $\alpha$ could be considerably improved either by increasing our sensitivity to substructures, i.e. by increasing the quality of the data, or by increasing the number of analysed objects. Although competing with the quality of HST seems at the moment difficult, future surveys such as LSST/JDEM in the optical and EVLA, e-MERLIN, LOFAR and SKA in the radio, will surely lead to an increase in the number of known lenses by several orders of magnitude \citep[see][]{Koopmans09,Marshall09} and dedicated optical and/or radio follow-up could provide equivalent or better data quality than HST. We expect therefore, in the foreseeable future to be able to characterise the galactic subhalo population with stringent constraint, both on the mass fraction and slope.

\item Although we have not explicitly performed a model comparison between different cosmologies, as for example CDM versus Warm Dark Matter (this would require an extra marginalization of the parameter space), the formalism introduced here, combined with the sensitivity of our method to CDM substructures, will allows us, in the future, to discriminate among these two scenarios and thus test the physics of dark matter.
\end{itemize}

\section*{Acknowledgements} 
SV and LVEK would like to thank Oliver Czoske and Phil Marshall for useful comments and feedback, and Mark Vogelsberger for providing details about the Aquarius simulation.
SV and LVEK are supported (in part) through an
NWO-VIDI program subsidy (project number 639.042.505).

\begin{table*}
\caption{Results for systems with no mass measurement error for a uniform (left) and a Gaussian (right) prior on $\alpha$. The input parameters are given in columns (1) to (3): the true mass fraction in substructure, $f_{\rm{true}}$, the lower detection mass threshold, $M_{\rm{low}}$, and the number of lens systems in a given sample, $n_l$. The maximum posterior values of $f$ and $\alpha$ are given in columns (4) and (5). Columns from (6) to (8) and (9) to (11) list the median, the 68\% and 95\% CL for $f$ and $\alpha$ respectively.} 
\label{tab:results}
\begin{center}
\begin{tabular}{ccccccccccc}
\hline
$f_{\rm{true}}$&$M_{\rm{low}}$&$n_l$&$f_{\rm{MP}}$&$\alpha_{\rm{MP}}$&$f_{\rm{med}}$&$\sigma_{f,68}$&$\sigma_{f,95}$&$\alpha_{\rm{med}}$&$\sigma_{\alpha,68}$&$\sigma_{\alpha,95}$\\
(\%)&$\left(10^{8}\msun\right)$&&(\%)&&(\%)&(\%)&(\%)\\
\hline
0.1&0.3&10&$0.14~|~0.14$&$1.91~|~1.90$&$0.18~|~0.15$&$^{-0.06}_{+0.11}~\big|~^{-0.04}_{+0.06}$&$^{-0.09}_{+0.22}~\big|~^{-0.07}_{+0.10}$&$1.88~|~1.90$&$^{-0.37}_{+0.48}~\big|~^{-0.05}_{+0.05}$&$^{-0.58}_{+0.81}~\big|~^{-0.09}_{+0.09}$\\
\\
&&30&$0.11~|~0.10$&$1.69~|~1.90$&$0.13~|~0.10$&$^{-0.04}_{+0.08}~\big|~^{-0.02}_{+0.03}$ &$^{-0.07}_{+0.15}~\big|~^{-0.04}_{+0.05}$&$1.64~|~1.90$&$^{-0.25}_{+0.29}~\big|~^{-0.03}_{+0.03}$ &$^{-0.40}_{+0.51}~\big|~^{-0.06}_{+0.05}$\\
\\
\\
&1.0&10&$0.55~|~0.16$&$2.70~|~1.90$&$0.60~|~0.20$&$^{-0.35}_{+0.70}~\big|~^{-0.10}_{+0.10}$ &$^{-0.45}_{+1.40}~\big|~^{-0.10}_{+0.20}$&$2.69~|~1.91$&$^{-0.57}_{+0.23}~\big|~^{-0.07}_{+0.07}$ &$^{-1.10}_{+0.29}~\big|~^{-0.11}_{+0.11}$\\
\\
&&30&$0.10~|~0.12$&$1.81~|~1.90$&$0.20~|~0.13$&$^{-0.09}_{+0.30}~\big|~^{-0.05}_{+0.06}$ &$^{-0.12}_{+0.69}~\big|~^{-0.07}_{+0.11}$&$2.14~|~1.90$&$^{-0.65}_{+0.60}~\big|~^{-0.04}_{+0.04}$ &$^{-0.93}_{+0.78}~\big|~^{-0.07}_{+0.07}$\\
\\
\\
&3.0&10&$1.53~|~0.10$&$2.77~|~1.90$&$2.15~|~0.25$&$^{-1.70}_{+4.06}~\big|~^{-0.15}_{+0.30}$ &$^{-2.00}_{+8.32}~\big|~^{-0.20}_{+0.55}$&$2.78~|~1.92$&$^{-0.46}_{+0.16}~\big|~^{-0.10}_{+0.10}$ &$^{-1.10}_{+0.20}~\big|~^{-0.16}_{+0.16}$\\
\\
&&30&$0.80~|~0.13$&$2.62~|~1.90$&$1.58~|~0.18$&$^{-1.19}_{+2.41}~\big|~^{-0.09}_{+0.13}$ &$^{-1.44}_{+4.67}~\big|~^{-0.13}_{+0.27}$&$2.78~|~1.91$&$^{-0.43}_{+0.16}~\big|~^{-0.07}_{+0.07}$ &$^{-1.01}_{+0.20}~\big|~^{-0.11}_{+0.12}$\\
\\
\\
0.5&0.3&10&$0.45~|~0.50$&$1.78~|~1.89$&$0.49~|~0.42$&$^{-0.12}_{+0.18}~\big|~^{-0.08}_{+0.09}$ &$^{-0.18}_{+0.34}~\big|~^{-0.13}_{+0.16}$&$1.76~|~1.90$&$^{-0.19}_{+0.21}~\big|~^{-0.03}_{+0.03}$ &$^{-0.31}_{+0.37}~\big|~^{-0.06}_{+0.05}$\\
\\
&&30&$0.49~|~0.50$&$1.90~|~1.90$&$0.51~|~0.50$&$^{-0.07}_{+0.09}~\big|~^{-0.06}_{+0.06}$ &$^{-0.11}_{+0.16}~\big|~^{-0.09}_{+0.11}$&$1.90~|~1.90$&$^{-0.12}_{+0.13}~\big|~^{-0.02}_{+0.01}$ &$^{-0.20}_{+0.23}~\big|~^{-0.03}_{+0.03}$\\
\\
&&200&$0.51~|~0.50$&$1.92~|~1.90$&$0.52~|~0.52$&$^{-0.05}_{+0.06}~\big|~^{-0.04}_{+0.04}$ &$^{-0.08}_{+0.10}~\big|~^{-0.07}_{+0.08}$&$1.92~|~1.90$&$^{-0.09}_{+0.09}~\big|~^{-0.01}_{+0.01}$ &$^{-0.15}_{+0.15}~\big|~^{-0.02}_{+0.02}$\\
\\
\\
&1.0&10&$0.52~|~0.48$&$2.04~|~1.90$&$0.85~|~0.52$&$^{-0.36}_{+1.28}~\big|~^{-0.15}_{+0.18}$ &$^{-0.49}_{+2.55}~\big|~^{-0.23}_{+0.32}$ &$2.33~|~1.90$&$^{-0.53}_{+0.44}~\big|~^{-0.04}_{+0.04}$ &$^{-0.83}_{+0.59}~\big|~^{-0.06}_{+0.06}$\\
\\
&&30&$0.42~|~0.50$&$1.68~|~1.90$&$0.47~|~0.44$&$^{-0.11}_{+0.14}~\big|~^{-0.10}_{+0.11}$ &$^{-0.16}_{+0.26}~\big|~^{-0.15}_{+0.19}$&$1.69~|~1.90$&$^{-0.26}_{+0.29}~\big|~^{-0.02}_{+0.02}$ &$^{-0.43}_{+0.51}~\big|~^{-0.04}_{+0.04}$\\
\\
&&200&$0.46~|~0.48$&$1.82~|~1.90$&$0.49~|~0.50$&$^{-0.07}_{+0.10}~\big|~^{-0.08}_{+0.09}$ &$^{-0.12}_{+0.18}~\big|~^{-0.13}_{+0.16}$&$1.84~|~1.90$&$^{-0.19}_{+0.20}~\big|~^{-0.02}_{+0.02}$ &$^{-0.30}_{+0.34}~\big|~^{-0.03}_{+0.03}$\\
\\
\\
&3.0&10&$1.85~|~0.46$&$2.56~|~1.90$&$5.91~|~0.60$&$^{-4.01}_{+6.92}~\big|~^{-0.30}_{+0.30}$ &$^{-5.21}_{+1.28}~\big|~^{-0.40}_{+0.60}$&$2.82~|~1.91$&$^{-0.31}_{+0.13}~\big|~^{-0.05}_{+0.06}$ &$^{-0.70}_{+0.16}~\big|~^{-0.09}_{+0.09}$\\
\\
&&30&$0.42~|~0.50$&$1.68~|~1.90$&$0.47~|~0.44$&$^{-0.11}_{+0.14}~\big|~^{-0.10}_{+0.11}$ &$^{-0.16}_{+0.26}~\big|~^{-0.15}_{+0.19}$&$1.69~|~1.90$&$^{-0.26}_{+0.29}~\big|~^{-0.02}_{+0.02}$ &$^{-0.43}_{+0.51}~\big|~^{-0.04}_{+0.04}$\\
\\
&&200&$0.50~|~0.50$&$1.90~|~1.90$& $0.35~|~0.43$&$^{-0.09}_{+0.13}~\big|~^{-0.12}_{+0.16}$&$^{-0.13}_{+0.30}~\big|~^{-0.19}_{+0.28}$&$1.35~\big|~1.90$&$^{-0.24}_{+0.38}~\big|~^{-0.03}_{+0.03}$ & $^{-0.32}_{+0.70}~\big|~^{-0.06}_{+0.05}$\\
\\
\\
2.5&0.3&10&$2.50~|~2.56$&$1.88~|~1.89$&$2.68~|~2.57$&$^{-0.27}_{+0.33}~\big|~^{-0.21}_{+0.22}$ &$^{-0.44}_{+0.59}~\big|~^{-0.34}_{+0.39}$&$1.85~|~1.89$&$^{-0.08}_{+0.08}~\big|~^{-0.03}_{+0.03}$ &$^{-0.13}_{+0.14}~\big|~^{-0.04}_{+0.05}$\\
\\
&&30&$2.39~|~2.31$&$1.82~|~1.89$&$2.42~|~2.33$&$^{-0.19}_{+0.21}~\big|~^{-0.16}_{+0.16}$ &$^{-0.31}_{+0.36}~\big|~^{-0.27}_{+0.28}$&$1.82~|~1.90$&$^{-0.07}_{+0.07}~\big|~^{-0.02}_{+0.01}$ &$^{-0.12}_{+0.13}~\big|~^{-0.03}_{+0.03}$\\
\\
\\
&1.0&10&$2.42~|~2.46$&$1.75~|~1.89$&$2.54~|~2.50$&$^{-0.36}_{+0.42}~\big|~^{-0.34}_{+0.39}$ &$^{-0.57}_{+0.74}~\big|~^{-0.56}_{+0.66}$&$1.76~|~1.90$&$^{-0.16}_{+0.17}~\big|~^{-0.03}_{+0.03}$ &$^{-0.27}_{+0.30}~\big|~^{-0.05}_{+0.05}$\\
\\
&&30&$2.12~|~2.14$&$1.70~|~1.89$&$2.16~|~2.15$&$^{-0.22}_{+0.25}~\big|~^{-0.21}_{+0.23}$ &$^{-0.35}_{+0.43}~\big|~^{-0.35}_{+0.39}$&$1.70~|~1.90$&$^{-0.11}_{+0.12}~\big|~^{-0.02}_{+0.01}$ &$^{-0.18}_{+0.20}~\big|~^{-0.03}_{+0.03}$\\
\\
\\
&3.0&10&$2.74~|~2.76$&$1.89~|~1.90$&$5.26~|~2.88$&$^{-2.76}_{+14.3}~\big|~^{-0.63}_{+0.75}$ &$^{-3.38}_{+35.8}~\big|~^{-1.00}_{+1.25}$&$2.22~|~1.90$&$^{-0.46}_{+0.44}~\big|~^{-0.03}_{+0.03}$ &$^{-0.73}_{+0.65}~\big|~^{-0.05}_{+0.05}$\\
\\  
&&30&$1.86~|~2.48$&$1.28~|~1.89$&$1.96~|~2.53$&$^{-0.32}_{+0.32}~\big|~^{-0.40}_{+0.44}$ &$^{-0.48}_{+0.60}~\big|~^{-0.60}_{+0.76}$&$1.31~|~1.90$&$^{-0.19}_{+0.21}~\big|~^{-0.02}_{+0.02}$ &$^{-0.27}_{+0.36}~\big|~^{-0.03}_{+0.03}$\\
\\
\hline
\end{tabular} 
\end{center}
\end{table*}

\newpage

\begin{table*}
\begin{center}
\caption{Results for systems with a mass measurement error of $0.1\times 10^{8}\msun$, $0.3\times 10^{8}\msun$, and $1.0\times 10^{8}\msun$ for a uniform (left) and a Gaussian (right) prior on $\alpha$. The input parameters are given in columns (1) to (3): the true mass fraction in substructure, $f_{\rm{true}}$, the lower detection mass threshold, $M_{\rm{low}}$, and the number of lens systems in a given sample, $n_l$. The maximum posterior values of $f$ and $\alpha$ are given in columns (4) and (5). Columns from (6) to (8) and (9) to (11) list the median, the 68\% and 95\% CL for $f$ and $\alpha$ respectively.}
\label{tab:results}
\begin{tabular}{cccccccccccc}
\hline
$f_{\rm{true}}$&$M_{\rm{low}}$&$n_l$&$f_{\rm{MP}}$&$\alpha_{\rm{MP}}$&$f_{\rm{med}}$&$\sigma_{f,68}$&$\sigma_{f,95}$&$\alpha_{\rm{med}}$&$\sigma_{\alpha,68}$&$\sigma_{\alpha,95}$\\
(\%)&$\left(10^{8}\msun\right)$&&(\%)&&(\%)&(\%)&(\%)\\
\hline
0.5&0.3&10&$0.53~|~0.61$&$2.24~|~1.91$&$0.57~|~0.64$&$^{-0.09}_{+0.09}~\big|~^{-0.12}_{+0.09}$ &$^{-0.15}_{+0.18}~\big|~^{-0.15}_{+0.18}$ &$2.25~|~1.91$&$^{-0.20}_{+0.22}~\big|~^{-0.02}_{+0.04}$ &$^{-0.32}_{+0.36}~\big|~^{-0.04}_{+0.06}$\\
\\
&&30&$0.53~|~0.53$&$1.91~|~1.90$&$0.54~|~0.54$&$^{-0.09}_{+0.09}~\big|~^{-0.06}_{+0.06}$ &$^{-0.12}_{+0.15}~\big|~^{-0.09}_{+0.09}$&$1.91~|~1.91$&$^{-0.12}_{+0.12}~\big|~^{-0.01}_{+0.01}$ &$^{-0.18}_{+0.20}~\big|~^{-0.04}_{+0.01}$\\
\\
&&200&$0.51~|~0.52$&$1.93~|~1.90$&$0.51~|~0.51$&$^{-0.06}_{+0.06}~\big|~^{-0.03}_{+0.06}$ &$^{-0.09}_{+0.12}~\big|~^{-0.06}_{+0.09}$ &$1.93~|~1.91$&$^{-0.08}_{+0.08}~\big|~^{-0.01}_{+0.01}$ &$^{-0.14}_{+0.14}~\big|~^{-0.02}_{+0.01}$\
\\
\\
&1.0&10&$0.53~|~0.44$&$2.18~|~1.90$&$0.82~|~0.48$&$^{-0.36}_{+0.94}~\big|~^{-0.15}_{+0.15}$&$^{-0.48}_{+1.79}~\big|~^{-0.21}_{+0.27}$& $2.41~|~1.91$&$^{-0.46}_{+0.38}~\big|~^{-0.04}_{+0.04}$ &$^{-0.77}_{+0.52}~\big|~^{-0.08}_{+0.06}$\\
\\
&&30&$0.55~|~0.54$&$1.98~|~1.90$&$0.64~|~0.54$&$^{-0.15}_{+0.18}~\big|~^{-0.09}_{+0.12}$ &$^{-0.21}_{+0.48}~\big|~^{-0.15}_{+0.21}$& $2.03~|~1.91$&$^{-0.24}_{+0.26}~\big|~^{-0.04}_{+0.02}$ &$^{-0.40}_{+0.44}~\big|~^{-0.04}_{+0.04}$\\
\\
&&200&$0.67~|~0.59$&$2.12~|~1.90$&$0.72~|~0.61$&$^{-0.15}_{+0.18}~\big|~^{-0.09}_{+0.09}$ &$^{-0.21}_{+0.39}~\big|~^{-0.12}_{+0.15}$ &$2.15~|~1.91$&$^{-0.16}_{+0.18}~\big|~^{-0.01}_{+0.01}$ &$^{-0.28}_{+0.30}~\big|~^{-0.04}_{+0.02}$\\
\\
&3.0&10&$4.69~|~0.42$& $2.85~|~1.90$& $4.01~|~0.50$& $^{-2.76}_{+4.76}~\big|~^{-0.25}_{+0.25}$ & $^{-3.51}_{+9.02}~\big|~^{-0.25}_{+0.50}$& $2.78~|~1.91$& $^{-0.38}_{+0.16}~\big|~^{-0.05}_{+0.06}$ & $^{-0.83}_{+0.20}~\big|~^{-0.09}_{+0.09}$\\ 
\\
&&30&$2.76~|~0.51$& $2.57~|~1.90$& $5.26~|~0.75$& $^{-3.51}_{+4.76}~\big|~^{-0.25}_{+0.1e-4}$&$^{-4.26}_{+8.02}~\big|~^{-0.25}_{+0.25}$& $2.76~|~1.91$&$^{-0.32}_{+0.17}~\big|~^{-0.04}_{+0.03}$ & $^{-0.62}_{+0.21}~\big|~^{-0.06}_{+0.06}$\\ 
\\
&&200&$0.51~|~0.51$& $1.90~|~1.90$& $3.76~|~0.75$& $^{-2.25}_{+4.26}~|~^{-0.25}_{+0.1e-4}$ & $^{-3.01}_{+7.02}~|~^{-0.25}_{+0.25}$& $2.68~|~1.90$& $^{-0.33}_{+0.22}~|~^{-0.03}_{+0.03}$ & $^{-0.58}_{+0.28}~|~^{-0.04}_{+0.05}$\\ 
\hline
\end{tabular} 
\end{center}
\end{table*}

\bibliography{ms}

\begin{thebibliography}{}

\bibitem[\protect\citeauthoryear{{Belokurov}, {Walker}, {Evans}
  et~al.,}{{Belokurov} et~al.}{2008}]{Belokurov08}
{Belokurov} V.,  {Walker} M.~G.,  {Evans} N.~W.,    et~al., 2008, \apjl, 686,
  L83

\bibitem[\protect\citeauthoryear{{Belokurov}, {Zucker}, {Evans}
  et~al.,}{{Belokurov} et~al.}{2006}]{Belokurov06}
{Belokurov} V.,  {Zucker} D.~B.,  {Evans} N.~W.,    et~al., 2006, \apjl, 647,
  L111

\bibitem[\protect\citeauthoryear{{Belokurov}, {Zucker}, {Evans}
  et~al.,}{{Belokurov} et~al.}{2007}]{Belokurov07b}
{Belokurov} V.,  {Zucker} D.~B.,  {Evans} N.~W.,    et~al., 2007, \apj, 654,
  897

\bibitem[\protect\citeauthoryear{Bergstr\"om, Edsj\"o, Gondolo \&
  Ullio}{Bergstr\"om et~al.}{1999}]{Bergstrom99}
Bergstr\"om L.,  Edsj\"o J.,  Gondolo P.,    Ullio P.,  1999, Phys. Rev. D, 59,
  043506

\bibitem[\protect\citeauthoryear{{Bolton}, {Burles}, {Koopmans}, {Treu} \&
  {Moustakas}}{{Bolton} et~al.}{2006}]{Bolton06}
{Bolton} A.~S.,  {Burles} S.,  {Koopmans} L.~V.~E.,  {Treu} T.,    {Moustakas}
  L.~A.,  2006, \apj, 638, 703

\bibitem[\protect\citeauthoryear{{Bolton}, {Treu}, {Koopmans}, {Gavazzi},
  {Moustakas}, {Burles}, {Schlegel} \& {Wayth}}{{Bolton}
  et~al.}{2008}]{Bolton08}
{Bolton} A.~S.,  {Treu} T.,  {Koopmans} L.~V.~E.,  {Gavazzi} R.,  {Moustakas}
  L.~A.,  {Burles} S.,  {Schlegel} D.~J.,    {Wayth} R.,  2008, \apj, 684, 248

\bibitem[\protect\citeauthoryear{Calc\'aneo-Rold\'an \&
  Moore}{Calc\'aneo-Rold\'an \& Moore}{2000}]{Calcaneo00}
Calc\'aneo-Rold\'an C.,  Moore B.,  2000, Phys. Rev. D, 62, 123005

\bibitem[\protect\citeauthoryear{Colafrancesco, Profumo \& Ullio}{Colafrancesco
  et~al.}{2006}]{Colafrancesco06}
Colafrancesco S.,  Profumo S.,    Ullio P.,  2006, \aap, 455, 21

\bibitem[\protect\citeauthoryear{{Czoske}, {Barnab{\`e}}, {Koopmans}, {Treu} \&
  {Bolton}}{{Czoske} et~al.}{2008}]{Czoske08}
{Czoske} O.,  {Barnab{\`e}} M.,  {Koopmans} L.~V.~E.,  {Treu} T.,    {Bolton}
  A.~S.,  2008, \mnras, 384, 987

\bibitem[\protect\citeauthoryear{Dalal \& Kochanek}{Dalal \&
  Kochanek}{2002}]{Dalal02}
Dalal N.,  Kochanek C.~S.,  2002, \apj, 572, 25

\bibitem[\protect\citeauthoryear{Diemand, Kuhlen \& Madau}{Diemand
  et~al.}{2007a}]{Diemand07b}
Diemand J.,  Kuhlen M.,    Madau P.,  2007a, \apj, 667, 859

\bibitem[\protect\citeauthoryear{Diemand, Kuhlen \& Madau}{Diemand
  et~al.}{2007b}]{Diemand07a}
Diemand J.,  Kuhlen M.,    Madau P.,  2007b, \apj, 657, 262

\bibitem[\protect\citeauthoryear{{Diemand}, {Kuhlen}, {Madau}, {Zemp}, {Moore},
  {Potter} \& {Stadel}}{{Diemand} et~al.}{2008}]{Diemand08}
{Diemand} J.,  {Kuhlen} M.,  {Madau} P.,  {Zemp} M.,  {Moore} B.,  {Potter} D.,
     {Stadel} J.,  2008, \nat, 454, 735

\bibitem[\protect\citeauthoryear{{Gao}, {White}, {Jenkins}, {Stoehr} \&
  {Springel}}{{Gao} et~al.}{2004}]{Gao04}
{Gao} L.,  {White} S.~D.~M.,  {Jenkins} A.,  {Stoehr} F.,    {Springel} V.,
  2004, \mnras, 355, 819

\bibitem[\protect\citeauthoryear{{Gavazzi}, {Treu}, {Koopmans}, {Bolton},
  {Moustakas}, {Burles} \& {Marshall}}{{Gavazzi} et~al.}{2008}]{Gavazzi08}
{Gavazzi} R.,  {Treu} T.,  {Koopmans} L.~V.~E.,  {Bolton} A.~S.,  {Moustakas}
  L.~A.,  {Burles} S.,    {Marshall} P.~J.,  2008, \apj, 677, 1046

\bibitem[\protect\citeauthoryear{{Gerhard}, {Kronawitter}, {Saglia} \&
  {Bender}}{{Gerhard} et~al.}{2001}]{Gerhard01}
{Gerhard} O.,  {Kronawitter} A.,  {Saglia} R.~P.,    {Bender} R.,  2001, \aj,
  121, 1936

\bibitem[\protect\citeauthoryear{{Grillmair}}{{Grillmair}}{2006}]{Grillmair06}
{Grillmair} C.~J.,  2006, \apjl, 645, L37

\bibitem[\protect\citeauthoryear{{Helmi}, {White} \& {Springel}}{{Helmi}
  et~al.}{2002}]{Helmi02}
{Helmi} A.,  {White} S.~D.,    {Springel} V.,  2002, \prd, 66, 063502

\bibitem[\protect\citeauthoryear{{Ibata}, {Martin}, {Irwin}, {Chapman},
  {Ferguson}, {Lewis} \& {McConnachie}}{{Ibata} et~al.}{2007}]{Ibata07}
{Ibata} R.,  {Martin} N.~F.,  {Irwin} M.,  {Chapman} S.,  {Ferguson} A.~M.~N.,
  {Lewis} G.~F.,    {McConnachie} A.~W.,  2007, \apj, 671, 1591

\bibitem[\protect\citeauthoryear{Ibata, Lewis, Irwin \& Quinn}{Ibata
  et~al.}{2002}]{Ibata02}
Ibata R.~A.,  Lewis G.~F.,  Irwin M.~J.,    Quinn T.,  2002, \mnras, 332, 915

\bibitem[\protect\citeauthoryear{{Irwin}, {Belokurov}, {Evans} et~al.,}{{Irwin}
  et~al.}{2007}]{Irwin07}
{Irwin} M.~J.,  {Belokurov} V.,  {Evans} N.~W.,    et~al., 2007, \apjl, 656,
  L13

\bibitem[\protect\citeauthoryear{{Jaffe}}{{Jaffe}}{1983}]{Jaffe83}
{Jaffe} W.,  1983, \mnras, 202, 995

\bibitem[\protect\citeauthoryear{Koopmans}{Koopmans}{2005}]{Koopmans05}
Koopmans L. V.~E.,  2005, \mnras, 363, 1136

\bibitem[\protect\citeauthoryear{{Koopmans}, {Bolton}, {Treu}
  et~al.,}{{Koopmans} et~al.}{2009}]{Koopmans209}
{Koopmans} L.~V.~E.,  {Bolton} A.,  {Treu} T.,    et~al., 2009, \apjl, 703, L51

\bibitem[\protect\citeauthoryear{{Koopmans} et~al.,}{{Koopmans}
  et~al.}{2009}]{Koopmans09}
{Koopmans} L.~V.~E.,  et~al., 2009, arXiv:0902.3186v2

\bibitem[\protect\citeauthoryear{{Koopmans} \& {Treu}}{{Koopmans} \&
  {Treu}}{2002}]{Koopmans02}
{Koopmans} L.~V.~E.,  {Treu} T.,  2002, \apjl, 568, L5

\bibitem[\protect\citeauthoryear{Koopmans, Treu, Bolton, Burles \&
  Moustakas}{Koopmans et~al.}{2006}]{Koopmans06}
Koopmans L. V.~E.,  Treu T.,  Bolton A.~S.,  Burles S.,    Moustakas L.~A.,
  2006, \apj, 649, 599

\bibitem[\protect\citeauthoryear{{Macci{\`o}} \& {Miranda}}{{Macci{\`o}} \&
  {Miranda}}{2006}]{Maccio06}
{Macci{\`o}} A.~V.,  {Miranda} M.,  2006, \mnras, 368, 599

\bibitem[\protect\citeauthoryear{MacKay}{MacKay}{1992}]{MacKay92}
MacKay D. J.~C.,  1992, PhD thesis, Caltech

\bibitem[\protect\citeauthoryear{{Majewski}, {Beaton}, {Patterson}
  et~al.,}{{Majewski} et~al.}{2007}]{Majewski07}
{Majewski} S.~R.,  {Beaton} R.~L.,  {Patterson} R.~J.,    et~al., 2007, \apjl,
  670, L9

\bibitem[\protect\citeauthoryear{Mao, Jing, Ostriker \& Weller}{Mao
  et~al.}{2004}]{Mao04}
Mao S.,  Jing Y.,  Ostriker J.~P.,    Weller J.,  2004, \apj, 604, L5

\bibitem[\protect\citeauthoryear{Mao \& Schneider}{Mao \&
  Schneider}{1998}]{Mao98}
Mao S.,  Schneider P.,  1998, \mnras, 295, 587

\bibitem[\protect\citeauthoryear{{Marshall}, {Auger}, {Bartlett}
  et~al.,}{{Marshall} et~al.}{2009}]{Marshall09}
{Marshall} P.~J.,  {Auger} M.,  {Bartlett} J.~G.,    et~al., 2009,
  arXiv:0902.2963, 2010, 194

\bibitem[\protect\citeauthoryear{{Martin}, {Ibata}, {Irwin}, {Chapman},
  {Lewis}, {Ferguson}, {Tanvir} \& {McConnachie}}{{Martin}
  et~al.}{2006}]{Martin06}
{Martin} N.~F.,  {Ibata} R.~A.,  {Irwin} M.~J.,  {Chapman} S.,  {Lewis} G.~F.,
  {Ferguson} A.~M.~N.,  {Tanvir} N.,    {McConnachie} A.~W.,  2006, \mnras,
  371, 1983

\bibitem[\protect\citeauthoryear{Mayer, Moore, Quinn, Governato \&
  Stadel}{Mayer et~al.}{2002}]{Mayer02}
Mayer L.,  Moore B.,  Quinn T.,  Governato F.,    Stadel J.,  2002, \mnras,
  336, 119

\bibitem[\protect\citeauthoryear{{Metcalf} \& {Madau}}{{Metcalf} \&
  {Madau}}{2001}]{Metcalf01}
{Metcalf} R.~B.,  {Madau} P.,  2001, \apj, 563, 9

\bibitem[\protect\citeauthoryear{{Sakamoto} \& {Hasegawa}}{{Sakamoto} \&
  {Hasegawa}}{2006}]{Sakamoto06}
{Sakamoto} T.,  {Hasegawa} T.,  2006, \apjl, 653, L29

\bibitem[\protect\citeauthoryear{{Schechter} \& {Wambsganss}}{{Schechter} \&
  {Wambsganss}}{2002}]{Schechter02}
{Schechter} P.~L.,  {Wambsganss} J.,  2002, \apj, 580, 685

\bibitem[\protect\citeauthoryear{{Springel}, {Wang}, {Vogelsberger}
  et~al.,}{{Springel} et~al.}{2008}]{Springel08}
{Springel} V.,  {Wang} J.,  {Vogelsberger} M.,    et~al., 2008, \mnras, 391,
  1685

\bibitem[\protect\citeauthoryear{{Stoehr}, {White}, {Springel}, {Tormen} \&
  {Yoshida}}{{Stoehr} et~al.}{2003}]{Stoehr03}
{Stoehr} F.,  {White} S.~D.~M.,  {Springel} V.,  {Tormen} G.,    {Yoshida} N.,
  2003, \mnras, 345, 1313

\bibitem[\protect\citeauthoryear{{Treu}, {Gavazzi}, {Gorecki}, {Marshall},
  {Koopmans}, {Bolton}, {Moustakas} \& {Burles}}{{Treu} et~al.}{2009}]{Treu09}
{Treu} T.,  {Gavazzi} R.,  {Gorecki} A.,  {Marshall} P.~J.,  {Koopmans}
  L.~V.~E.,  {Bolton} A.~S.,  {Moustakas} L.~A.,    {Burles} S.,  2009, \apj,
  690, 670

\bibitem[\protect\citeauthoryear{{Vegetti} \& {Koopmans}}{{Vegetti} \&
  {Koopmans}}{2009}]{Vegetti09}
{Vegetti} S.,  {Koopmans} L.~V.~E.,  2009, \mnras, 392, 945

\bibitem[\protect\citeauthoryear{{Walsh}, {Jerjen} \& {Willman}}{{Walsh}
  et~al.}{2007}]{Walsh07}
{Walsh} S.~M.,  {Jerjen} H.,    {Willman} B.,  2007, \apjl, 662, L83

\bibitem[\protect\citeauthoryear{{Willman}, {Dalcanton}, {Martinez-Delgado}
  et~al.,}{{Willman} et~al.}{2005}]{Willman05}
{Willman} B.,  {Dalcanton} J.~J.,  {Martinez-Delgado} D.,    et~al., 2005,
  \apjl, 626, L85

\bibitem[\protect\citeauthoryear{{Zucker}, {Belokurov}, {Evans}
  et~al.,}{{Zucker} et~al.}{2006a}]{Zucker06b}
{Zucker} D.~B.,  {Belokurov} V.,  {Evans} N.~W.,    et~al., 2006a, \apjl, 650,
  L41

\bibitem[\protect\citeauthoryear{{Zucker}, {Belokurov}, {Evans}
  et~al.,}{{Zucker} et~al.}{2006b}]{Zucker06a}
{Zucker} D.~B.,  {Belokurov} V.,  {Evans} N.~W.,    et~al., 2006b, \apjl, 643,
  L103

\bibitem[\protect\citeauthoryear{{Zucker}, {Kniazev}, {Bell} et~al.,}{{Zucker}
  et~al.}{2004}]{Zucker04}
{Zucker} D.~B.,  {Kniazev} A.~Y.,  {Bell} E.~F.,    et~al., 2004, \apjl, 612,
  L121

\bibitem[\protect\citeauthoryear{{Zucker}, {Kniazev}, {Mart{\'{\i}}nez-Delgado}
  et~al.,}{{Zucker} et~al.}{2007}]{Zucker07}
{Zucker} D.~B.,  {Kniazev} A.~Y.,  {Mart{\'{\i}}nez-Delgado}   et~al., 2007,
  \apjl, 659, L21

\end{thebibliography}
\topmargin -1.3cm
\newpage
\appendix
\section{}
We show here that the procedure of first drawing objects between $M_{\rm{min}}$ and $M_{\rm{max}}$ from $dP/dm|_{true}$, then scattering with a Gaussian and finally restricting to those masses between $M_{\rm{low}}$ and $M_{\rm{high}}$ gives a probability $P(N_{\rm{obs}})$ of observing $N_{\rm{obs}}$ objects that is equivalent to the Poisson probability density distribution, with expectation value $\mu_s$, of $N_{\rm{obs}}$ objects between $M_{\rm{low}}$ and $M_{\rm{high}}$ expressed by the convolution in equation (\ref{equ:mu_conv}). \\
\noi Let us divide the mass ranges $\left[M_{\rm{min}}, M_{\rm{max}}\right]$ and $\left[M_{\rm{low}}, M_{\rm{high}}\right]$ respectively in $n$ and $n^\prime$ sub-intervals of infinitesimally small widths $dm$ and $dm^\prime$. Lets call $p_{s,j}$ the probability that an object is scattered from the $j$-th mass bin $dm_j$ to the $k$-th one $dm_k^\prime$.
In the particular case of Gaussian errors, $p_{s,j,k}$ reads as follows
\begin{equation}
p_{s,j,k} = \frac{e^{-\left(m-m^\prime\right)^2/2\sigma^2}}{\sqrt{2\pi}\sigma}~dm_k^{\prime}\,,
\label{equ:scatter}
\end{equation} 
\noi with $m\in dm_j$ and $m^\prime\in dm_k^\prime$.\\
\noi First we show that, if substructures are Poisson distributed in each mass bin with expectation value $d\mu_j$, then the probability of having $N_s$ objects scattered from $dm_j$ into $dm_k^{\prime}$ is also a Poisson distribution with expectation value $d\mu_{s,j,k}=p_{s,j,k}~d\mu_j$
\begin{multline}
P_{s,j}(N_s,d\mu_{s,j,k})= \sum_{i=N_s}^{\infty}{P(i,d\mu_j)~\binom {i}{N_s}~p_{s,j,k}^{N_s}~\left(1-p_{s,j,k}\right)^{i-N_s}}=\\  \sum_{i=N_s}^{\infty}{\frac{e^{-d\mu_j}~d\mu_j^i}{i!} ~\binom {i}{N_s}~p_{s,j,k}^{N_s}~\left(1-p_{s,j,k}\right)^{i-N_s}}=\\
\frac{p_{s,j,k}^{N_s}}{N_s!}\lim_{n\rightarrow\infty} \sum_{i=N_s}^{n}\frac{e^{-d\mu_j}~d\mu_j^i}{\left(i-N_s\right)!}\left(1-p_{s,j,k}\right)^{i-N_s}=\\ \left(\frac{p_{s,j,k}}{1-p_{s,j,k}}\right)^{{N_s}}\frac{e^{-d\mu_j}~d\mu_j^{N_s}}{N_s!}\lim_{n\rightarrow\infty} \sum_{k=0}^{n-N_s}\frac{d\mu_j^k}{k!}\left(1-p_{s,j,k}\right)^{k}=\\ \frac{e^{-p_{s,j,k}~\mu_j}~\left(p_{s,j,k}~d\mu_j\right)^{N_s}}{N_s!}\,,
\end{multline} \\              
\noi hence it follows that $d\mu_{s,j,k}=p_{s,j,k}~d\mu_j$. This result directly follows from the fact that in the case of high number statistics the Binomial tends to a Poisson distribution and from the product rule of the Poisson distribution. Specifically, each $d\mu_j$ reads as
\begin{equation}
d\mu_j= \mu_0(\alpha,f,R)~ \left.\frac{dP}{dm}\right |_{true}~dm_j\,.
\end{equation}
\noi We now extend this result to two mass intervals of the same size $dm$; thanks to the sum rule, the probability of $N_{obs}$ objects being scattered is again Poissonian with $d\mu_{s,k}=(p_{s,1,k}d\mu_1+p_{s,2,k}d\mu_{2})$
 \begin{multline}
 P(N_{obs},d\mu_s)=\sum_{i=0}^{N_{obs}}P_{s,1}(i) P_{s,2}(N_{obs}-i)=\\ \sum_{i=0}^{N_{obs}}\frac{e^{-p_{s,1,k}d\mu_1}\left(p_{s,1,k}~d\mu_1\right)^i}{i!} \frac{e^{-p_{s,2,k}d\mu_{2}}\left(p_{s,2,k}~d\mu_{2}\right)^{N_{obs}-i}}{\left(N_{obs}-i\right)!}=\\
 e^{\left(p_{s,1,k}d\mu_1+p_{s,2,k}d\mu_{2}\right)}\left(p_{s,2}~d\mu_{2}\right)^{N_{obs}}\sum_{i=0}^{N_{obs}}\frac{d\mu_1^i}{i!}\frac{d\mu_{2}^{-i}}{\left(N_{obs}-i\right)!}=\\
 e^{-\left(p_{s,1,k}d\mu_1+p_{s,2,k}d\mu_{2}\right)}\left(p_{s,2,k}~d\mu_{2}\right)^{N_{obs}}\frac{\left(1+\frac{p_{s,1,k}~d\mu_1}{p_{s,2,k}~d\mu_{2}}\right)^{N_{obs}}}{N_{obs}!}=\frac{e^{-d\mu_{s,k}}~ d\mu_{s,k}^{N_{obs}}}{N_{obs}!}\,.
 \end{multline}
\noi By induction it can be shown that in the case of a generic number $n$ of intervals $d\mu_{s,k}=\sum_{j=1}^np_{s,j,k}~d\mu_j$, so that the probability of being scattered outside  $\left[M_{\rm{min}}, M_{\rm{max}}\right]$ and inside $\left[M_{\rm{low}}, M_{\rm{high}}\right]$ is a Poisson distribution with expectation value 
\begin{equation}
\mu_s=\sum_{k=1}^{n^\prime}\sum_{j=1}^np_{s,j,k}~d\mu_j= \sum_{k=1}^{n^\prime}\sum_{j=1}^np_{s,j,k}~ \mu_0(\alpha,f,R)~ \left. \frac{dP}{dm}\right |_{true}~dm_j~dm_k^\prime\,.
\end{equation}
\noi In the limit of equally infinitesimal intervals, i.e $dm\rightarrow0$ and $dm^\prime\rightarrow0$ and making use of equation (\ref{equ:scatter}) this becomes
\begin{equation}
\mu_s=\mu_0(\alpha,f,R)~\int_{M_{\rm{low}}}^{M_{\rm{high}}}{\int_{M_{\rm{min}}}^{M_{\rm{max}}}\frac{e^{-\left(m-m^\prime\right)^2/2\sigma^2}}{\sqrt{2\pi}\sigma}~\left.\frac{dP}{dm}\right |_{true}~dm~dm^\prime}\,.
\end{equation}
\noi q.e.d
\clearpage
\end{document}